\documentclass[11pt]{article}

\usepackage[marginratio=1:1,height=600pt,width=460pt,tmargin=100pt]{geometry}

\pdfoutput = 1

\usepackage{authblk}
\usepackage{amsmath}
\usepackage{amsthm}
\usepackage{bm}
\usepackage{amsfonts}
\usepackage{amssymb}
\usepackage{color}
\usepackage{graphicx}
\usepackage[authoryear,sort]{natbib}
\usepackage{float}
\usepackage{wrapfig}
\usepackage[hidelinks]{hyperref}
\usepackage{siunitx}
\usepackage{caption}
\usepackage{subcaption}
\usepackage{mathtools}
\usepackage[english]{babel}
\usepackage[normalem]{ulem}

\setcitestyle{authoryear, round}

\newtheorem{prop}{Proposition}
\newtheorem*{remark}{Remark}

\newcommand\numberthis{\addtocounter{equation}{1}\tag{\theequation}}
\newcommand*\mystrut[1]{\vrule width0pt height0pt depth#1\relax}

\title{Non-stationary spatio-temporal point process modeling for high-resolution COVID-19 data}

\author[1]{Zheng Dong}
\author[2]{Shixiang Zhu}
\author[1]{Yao Xie}
\author[3]{Jorge Mateu}
\author[4]{Francisco J. Rodr\'iguez-Cort\'es}
\affil[1]{H. Milton Stewart School of Industrial and Systems Engineering, Georgia Institute of Technology, Atlanta, USA}
\affil[2]{Heinz College of Information Systems and Public Policy, Carnegie Mellon University, Pittsburgh, USA}
\affil[3]{Department of Mathematics, Universitat Jaume I, Castell\'o de la Plana, Valencia, Spain}
\affil[4]{Escuela de Estad\'istica, Universidad Nacional de Colombia, Medell\'in, Colombia}
\date{}

\allowdisplaybreaks

\begin{document}

\maketitle

\begin{abstract}

Most COVID-19 studies commonly report figures of the overall infection at a state- or county-level. This aggregation tends to  miss out on fine details of virus propagation. In this paper, we analyze a high-resolution COVID-19 dataset in Cali, Colombia, that records the precise time and location of every confirmed case. We develop a non-stationary spatio-temporal point process equipped with a neural network-based kernel to capture the heterogeneous correlations among COVID-19 cases. The kernel is carefully crafted to enhance expressiveness while maintaining model interpretability. We also incorporate some exogenous influences imposed by city landmarks. Our approach outperforms the state-of-the-art in forecasting new COVID-19 cases with the capability to offer vital insights into the spatio-temporal interaction between individuals concerning the disease spread in a metropolis.

\end{abstract}

\section{Introduction}

The outbreak of coronavirus disease 2019 (COVID-19) since 2020 has swept the world and is still developing. It causes a dramatic loss of human lives \citep{COVID-19_intro} and presents an unprecedented challenge to public health, food systems, and the world. 
Tracking the dynamics of COVID-19 enables human beings to take target-protecting measures to curb the pandemic's spread and design health surveillance systems. However, limited and biased information about local COVID-19 cases makes it extremely difficult to effectively control strategies against the pandemic.

There is a large amount of {\it aggregated} data consistently collected and publicly available, which contains rich information about COVID-19 cases. For instance, Johns Hopkins Center for Systems Science and Engineering (JHU CSSE) establishes an interactive COVID-19 dashboard to track the global coronavirus development \citep{COVID19JHU} which reports the daily confirmed cases and deaths worldwide up to the state level. \citet{COVID19NYTimes} also tracks daily the county-level counts of confirmed cases and deaths in the United States. Such data help the scientific researcher model the disease transmission on an aggregated level and play a pivotal role in tracking the propagation patterns of the virus and helping policymakers act effectively to revitalize economic and social development.

However, such aggregated data lack precise information about individual cases and present a significant challenge in modeling the spatio-temporal dynamics of human-to-human disease transmission when capturing the fine spatial heterogeneity of case distribution in a small region. 
Aggregated data may lose fine-grained spatio-temporal information, which will lead administrative officials to make biased decisions. For example, it is reported in \cite{trackCOVID-19Italy} that the unreliable preliminary data, as well as inaccurate models, significantly affected the political decisions of the Italian administration. Another example in \cite{superspreadinGermany} documents a superspreader event in Germany in a meat processing plant; modeling such an event requires accounting for the precise plant location information, and aggregated data may miss such crucial local information. 

In this paper, we consider a unique high-resolution dataset for individual cases of COVID-19 in Cali, Colombia, the second-largest city in the country. This data records individual confirmed cases for six months, from March 15 to September 30 of 2020, with each case's time and location information. To take full advantage of the fine-grained dataset, we develop a non-stationary spatio-temporal point process model, assuming that previously infected events trigger the newly confirmed cases. We assume the triggering effect is non-stationary \citep{hendry2016all} since the virus is likely to spread more slowly in sparsely populated rural than in densely populated areas. This fact entails stationary point processes non-applicable: the stationary kernel is ``shift-invariant'' and only depends on the temporal and location differences between events. Moreover, we consider the exogenous promotion of densely populated city landmarks in the model since the COVID-19 virus proves to spread quickly through respiratory droplets \citep{COVID19transLancet}, and aerosol transmission in crowded and inadequately ventilated spaces \citep{indoorinfection}. We represent parameters of the non-stationary kernel by neural networks to enhance model flexibility while maintaining interpretability. The model is estimated by solving a maximum likelihood problem via a computationally efficient strategy to tackle the intractable numerical integration in the log-likelihood function. We conduct an extensive real-data study, which reveals the unique transmission dynamics of COVID-19 and confirms that a few landmarks in the city play an essential role in spreading the virus.
The model and results will help policymakers monitor coronavirus dynamics and provide a template for tracking real-time data for future epidemics and implementing health surveillance systems. 
Since similar high-resolution datasets will not be so rare in the future, the need for such an approach is not limited to the situation of Cali.

The paper is organized as follows. The rest of this section discusses some relevant literature on COVID-19 modeling and spatio-temporal point processes. We then introduce our motivating (and unique) dataset in Section \ref{sec:data-description}. In Section \ref{sec:proposed-method}, we review basic knowledge about point processes, propose our framework with a non-stationary spatio-temporal kernel, and illustrate our fine-crafted parameterization scheme with a simple neural network. Section \ref{sec:efficient-learning} presents the computational strategies for model estimation with an approximation to the likelihood. Section \ref{sec:results} interprets the results from real data and compares them with several benchmark models. Lastly, Section \ref{sec:conclusion} concludes the paper.

\paragraph{Related work.} 
Compartmental models are widely developed to describe the overall COVID-19 infection in a region. The simplest SIR compartmental model \citep{SIR} assigns the population into three compartments with labels $\bm{S}$ (susceptible), $\bm{I}$ (infectious), and $\bm{R}$ (recovered), respectively. Deterministic differential equations fit the transition rates between each kind of compartment. 
Advanced compartmental models are further designed by reframing the basic one with different compartments \citep{ConceptualModel, DynSocialDistancing}.
SEIRD in \cite{SEIRDinCOVID19} and forced SEIRD models in \cite{fSEIRDinCOVID19} are adopted in various epidemic scenarios by introducing compartments of exposed and deceased populations into the system. Other extensions, such as splitting the infected population according to infection severity \citep{DynSocialDistancing} and introducing unreported infected population \citep{ConceptualModel} are also considered. Compartmental models assume a stable population of the inspected region, thus perform well when applied to large regions such as a country or state. However, they usually do not consider detailed spatial information, such as population migration across regions. 

Much work has been done on predicting the number of COVID-19 cases and deaths.
\cite{IntegerModel} and \cite{woody2020projections} adopted Generalized linear models to predict the number of daily cases and deaths during the first-wave of COVID-19 in China and the United States, respectively.
Autoregressive models are also widely used to forecast confirmed cases at a state-level \citep{ARwithCovariates, ARFirst-order, Agosto2020PoissonAR}. 
There are also several studies \citep{MOBS, 2020IHME-CDC} adopted by the Centers for Disease Control and Prevention (CDC) for COVID-19 case forecast in the United States. 
Our approach differs from these methods in two ways: (a) Our model provides finer-grained predictions based on the unique data, and (b) we focus on capturing the spatio-temporal correlation between confirmed cases and emphasize the interpretability of the proposed model.

Spatio-temporal analysis of COVID-19 plays a pivotal role in understanding the dynamics of the spread of COVID-19.
\cite{SpatialtemporalSIR} introduces a spatio-temporal BME-SIR model integrating the disease representation at different locations to generate disease predictions.
\cite{Bai2020NonstatSIR} divides the regional-level COVID-19 time series data in the United States into several periods and develops a piecewise stationary SIR model coupled with spatio-temporal dependence. 
In addition, a vector autoregressive model developed by \cite{zhu2021highresolution} considers local spatio-temporal correlations, mobility, and demographic factors, aiming to estimate COVID-19 cases and deaths at a county level in the United States.
In \cite{CovariatesHawkesModeling}, a multivariate Hawkes process is adopted to model the occurrence of confirmed cases across the U.S. counties by incorporating social and health covariates. 
However, most of these methods use spatially or temporally aggregated data, which hinders us from understanding the spread of COVID-19 at an individual level.

A few studies attempt to model the dynamics of COVID-19 using point processes. \cite{Gajardo2021PPforCOVID19} proposes a point process regression framework of COVID-19 cases and deaths conditioned on mobility and economic covariates. \cite{Giudici2021NetworkPP} focuses on country-level case prediction in 27 European countries by augmenting spatio-temporal point process model with mobility network covariates.
\cite{ReinformentSTPP} introduces a generative and intensity-free point process model based on an imitation learning framework to track the spread of COVID-19 and forecast county-level cases in the United States.  Compared to the previous methods, our approach is more flexible by considering non-stationarity in the spatial correlation, which is highly interpretable and expressive in representing the spread of the disease. 

Recent works \citep{berry2020open, branden2020residential, lopez2021air} investigate the relationship between various factors and COVID-19 by collecting confirmed COVID-19 cases and mortality data at the individual level. However, these data are aggregated and reported at the county or state level. Other works \citep{fu2020tocilizumab, guo2020tocilizumab} use individual patient data, including peripheral blood samples, observed monocytes, and T-cell data from patients with severe COVID-19 symptoms to demonstrate the effectiveness of COVID-19 treatments, which is different from our objective in this paper.

Last but not least, some related studies use similar techniques developed in this paper.
\cite{RMTPP, NeuralHawkes, AttentiveHawkes} model discrete events using neural-network-based point process models. However, most of these works aim to enhance the representative power by taking advantage of the recurrent neural structure \citep{Hochreiter1997LSTM} or the attention mechanism \citep{vaswani2017attention} to represent the historical information, which lacks interpretability and is unable to capture long-term effects. 
A wide array of research focuses on characterizing the triggering effects between events using a fine-crafted kernel function. 
Original works \cite{Ogata1988, Ogata1998} introduce a parametric kernel in Epidemic Type Aftershock Sequence (ETAS) to capture the triggering effects between earthquakes.
There are a few prior works in studying non-stationary kernels in Gaussian processes. Some studies \citep{Higdon1998NonStationarySM, lang2007adaptive, vasudevan2011non} aim to model spatial correlation based on a squared exponential kernel whose covariance structure is location-dependent. Another paper \citep{Remes2017NonStationarySK} develops a non-stationary spectral kernel based on a generalized Fourier transform, whose spectral density is represented as a Gaussian mixture. Another line of research focuses on developing non-stationary kernels in point processes \citep{zhu2021imitation, zhu2021deep, zhu2021early, zhu2022neural}. These studies capture intricate spatio-temporal dependence between discrete events by introducing neural networks to kernel designs. Our work is a significant extension of the previous approaches, which greatly enhances the expressiveness of the non-stationary kernel by considering both inter- and intra-influences between spatial kernel feature functions while still being highly interpretable and computationally efficient.
A few works incorporate the exogenous effects into point process models by adding terms in the conditional intensity function \citep{zhu2021spatio, Rizoiu_Exogenous_Media, farajtabar2017fake}.

\section{Data description and preliminary analysis}
\label{sec:data-description}

\begin{figure}[!t]
\centering 
\includegraphics[width=0.33\linewidth]{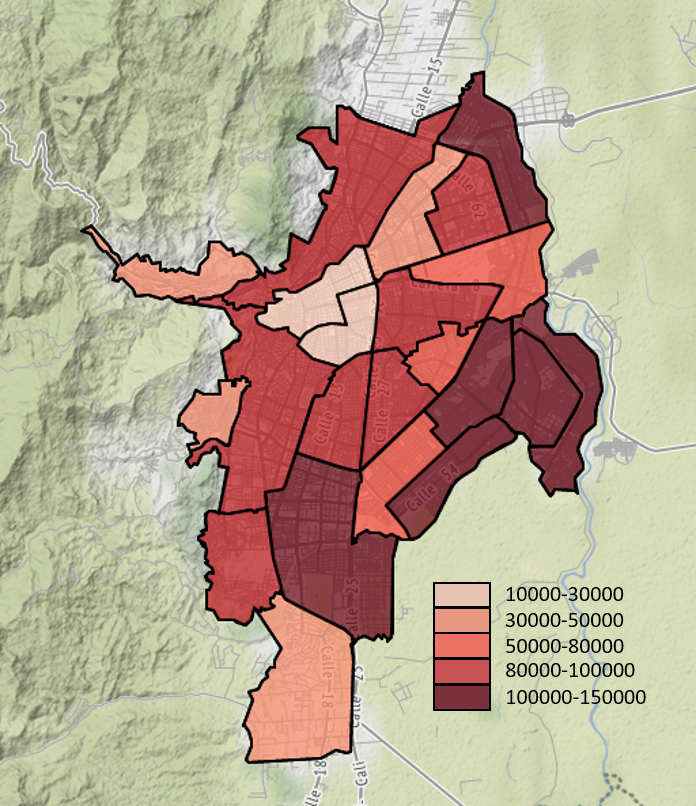}
\includegraphics[width=0.333\linewidth]{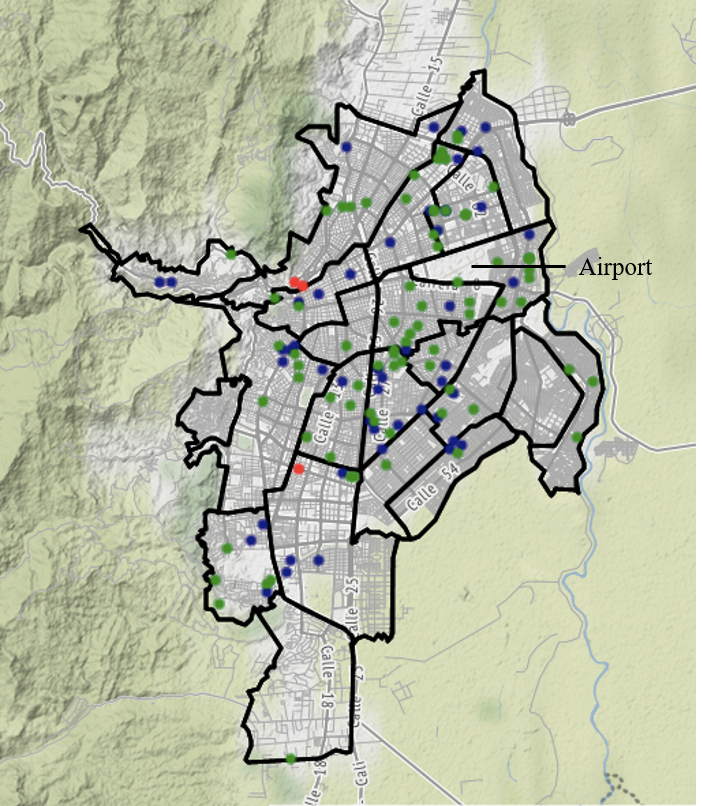}
\caption{\textit{Left panel}: Population distribution in Cali. 
Each polygon bounded by bold lines represents a comuna (a municipality-level subdivision in Cali); there are 22 comunas in the city of Cali. 
\textit{Right panel}: Landmarks in Cali. 
Each dot represents the landmark's location, and its color indicates the type of the landmark, where the red dot is a town hall, the blue dot is a church, and the green dot is a school. 
}
\label{fig:terrain-and-population}
\end{figure}

The COVID-19 dataset provided by the Municipal Public Health Secretary of Cali\footnote{\url{https://www.cali.gov.co/salud/}} documents the individual-level confirmed COVID-19 cases collected from Cali, one of the major cities in Colombia, the capital of the Valle del Cauca department and the most populated city in southwest Colombia, with 2,227,642 residents according to the 2018 census. 
As shown in Fig.~\ref{fig:terrain-and-population}(a), more than half of the population concentrates in neighborhoods of low socioeconomic strata located mainly in the east, northeast, and west. Almost a tenth of the population is under the line of poverty agglomerates in the city's eastern neighborhoods. The population with higher socioeconomic strata distributes in the other city areas, concentrating the wealthiest population in the city's south. The city spans 560.3 square kilometers (216.3 square miles) with 120.9 square kilometers (46.7 square miles) of urban area, making it the second-largest and the third-most populated city in the country. 
As the only major Colombian city with access to the Pacific coast, Cali is the leading industrial and economic center in the country's south, with one of Colombia's fastest-growing economies. Cali's international airport is located in the northeast part of the city, and it is Colombia's third-largest airport in terms of passengers \citep{Cali_intro}.

\begin{figure}[!t]
\centering 
\includegraphics[width=.24\linewidth]{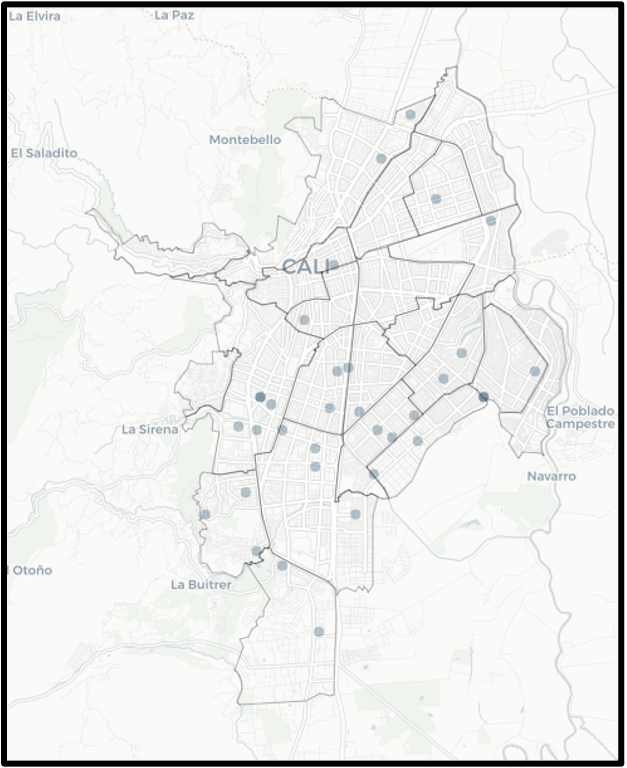}
\includegraphics[width=.24\linewidth]{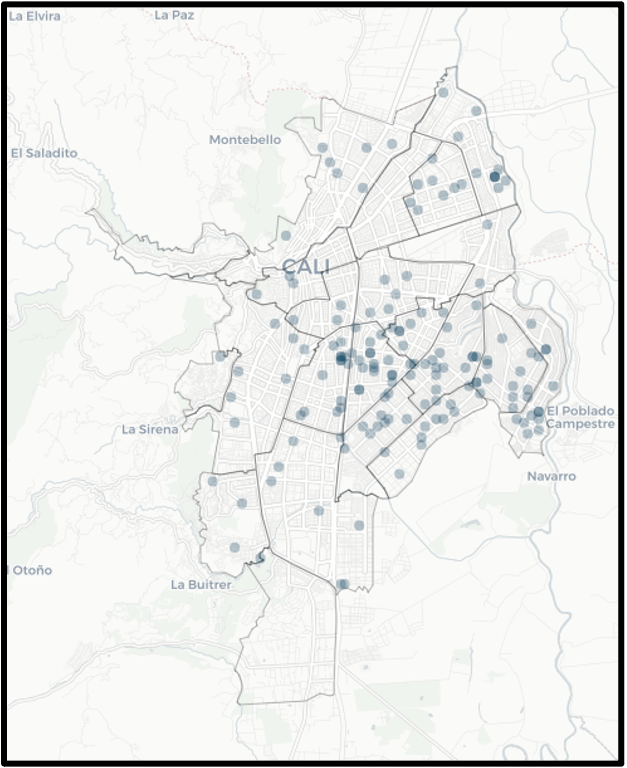}
\includegraphics[width=.24\linewidth]{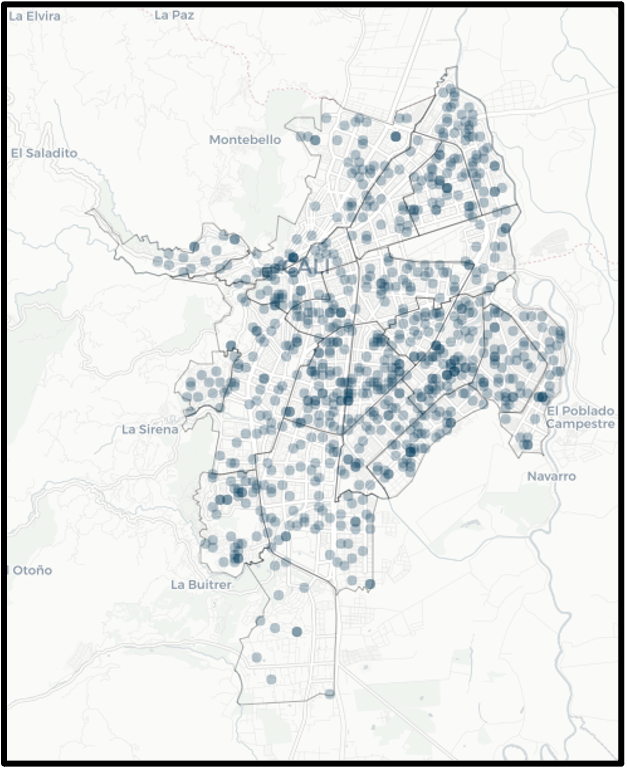}
\includegraphics[width=.24\linewidth]{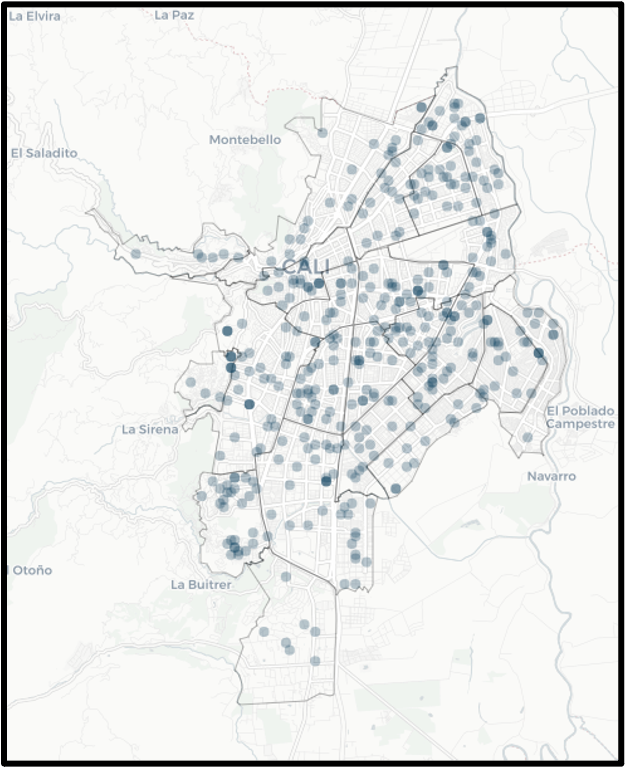}
\caption{Snapshots of confirmed COVID-19 cases at four particular weeks in 2020: March 29, May 17, July 12, and August 23. Each dot represents the location of a confirmed case. 
Note that darker dots indicate multiple dots being overlapped.}
\label{fig:cases-snapshots}
\end{figure}

The dataset records 38,611 cases from March 15 to September 30 of 2020, including 28 weeks. 
Specifically, a COVID-19 case was recorded once confirmed, with the diagnosed date of the patient and the geographical location (measured in longitude and latitude) of their residence. 
The testing procedures were carried out across the entire urban area, with similar testing rates in each comuna.
Unlike other commonly-seen COVID-19 datasets that only report the aggregated number of cases or deaths at a state or county level, this dataset records each confirmed case's exact location and time.
In practice, we observe periodic weekly oscillations in daily reported cases and deaths, which may be caused by testing bias (higher testing rates on certain days of the week). 
To reduce such bias, we aggregate the number of cases and deaths
of each county by weeks.
Fig.~\ref{fig:cases-snapshots} presents the spatial distribution of confirmed cases at four particular weeks in Cali.
We note that the first confirmed case of COVID-19 in Colombia appeared on March 6, 2020. On March 12, the country soon declared a state of emergency. 
On March 15, Cali reported the first positive person. 
Then the authorities announced the mandatory isolation of the entire city for just eight days \citep{COVID19COL}. The first case reported in the city based on people who went to health services occurred in high socioeconomic strata. However, the disease quickly spread and concentrated in the most vulnerable areas with low socioeconomic strata. After early efforts of the government to contain the pandemic, inevitably, the virus spread throughout the city, affecting a large part of the population. 
The above public health decisions are known not significantly to affect the dynamics of the virus spreading. Thus, we do not consider the impact of these decisions in our model for simplicity.


Besides COVID-19 events, we also collect the location of three kinds of landmarks in Cali, including churches, schools, and town halls, from the Administrative Department of Municipal Planning\footnote{\url{https://www.cali.gov.co/planeacion/}}.
These locations play an important role in understanding the wide and rapid spreading of the virus.
According to \cite{web2021high},
there is a high COVID-19 positive rate among attendees to events at places such as churches. As a clear note in this line, among 92 attendees at a rural Arkansas church during March 6–11, 35 (38\%) developed laboratory-confirmed COVID-19, and three persons died \citep{web2021high}. The landmark dataset has three town halls, 49 small and large churches, and 77 schools. 
Fig.~\ref{fig:terrain-and-population}(b) shows the exact locations of these collected landmarks. 



\begin{figure}[t]
\centering 
\includegraphics[width=.33\linewidth]{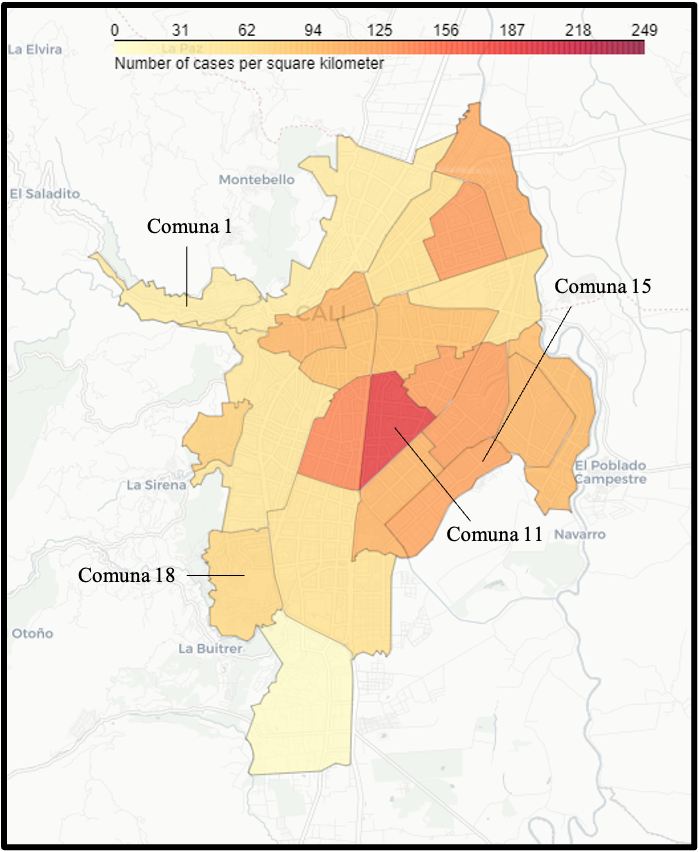}
\includegraphics[width=.33\linewidth]{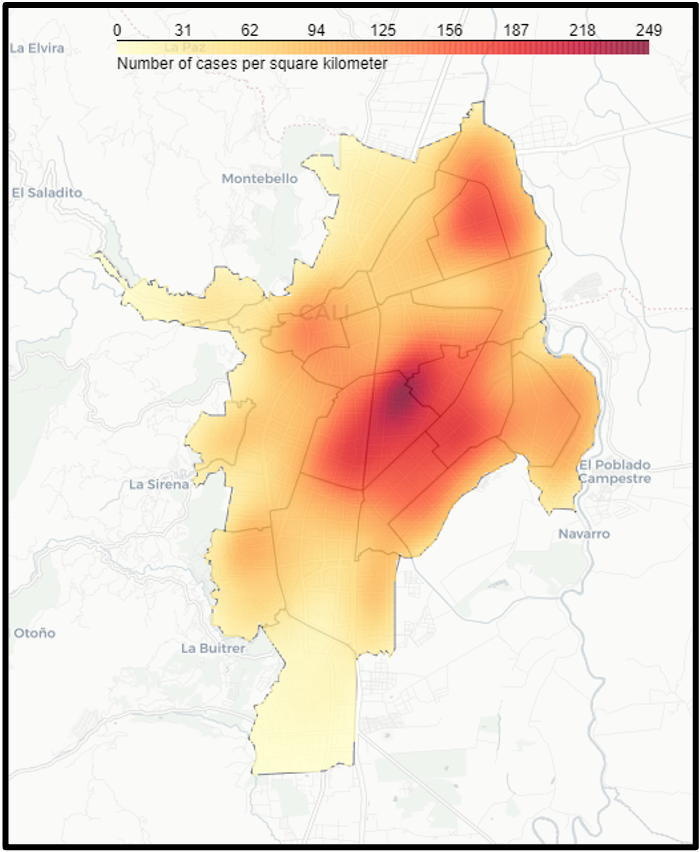}
\caption{Spatial distribution of the confirmed cases at two spatial resolutions. The color depth indicates the number of confirmed cases in one square kilometer; \textit{Left panel} shows the case density per comuna. \textit{Right panel} shows the spatially continuous case density estimated by KDE.}
\label{fig:resolutions}
\end{figure}

\begin{figure}[t]
\centering 
\includegraphics[width=.24\linewidth]{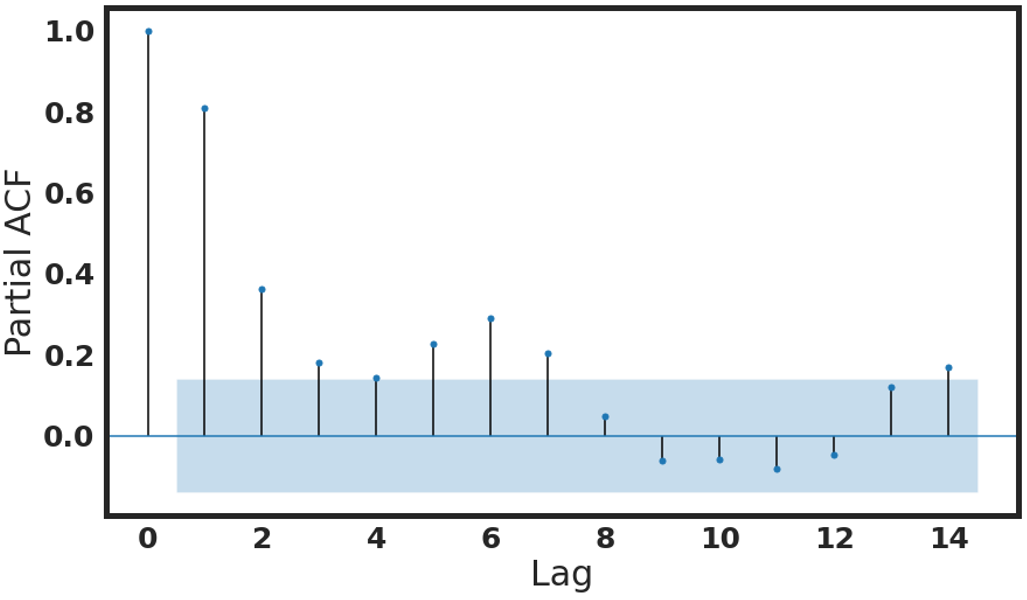}
\includegraphics[width=.24\linewidth]{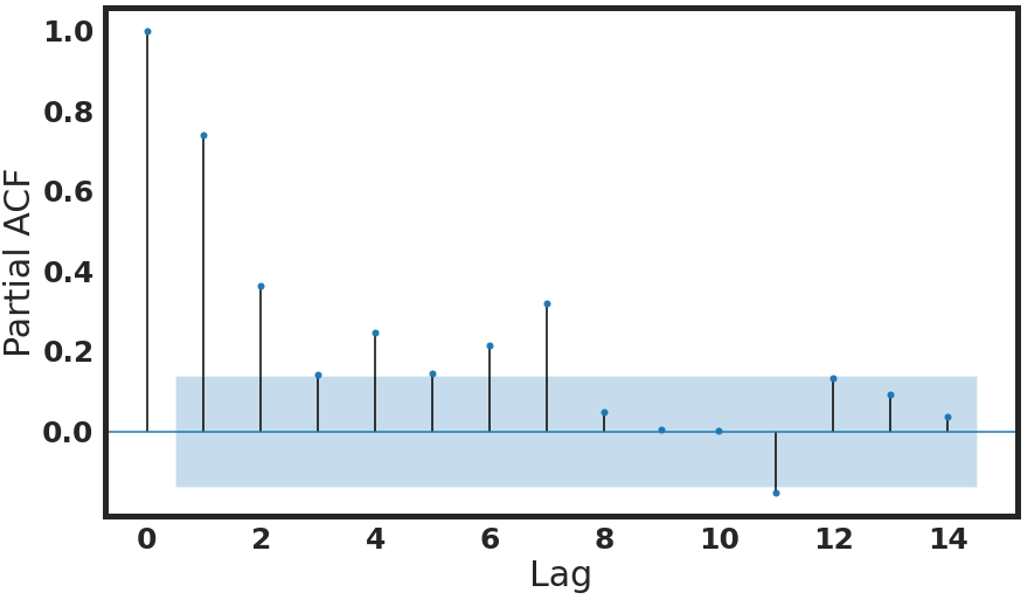}
\includegraphics[width=.24\linewidth]{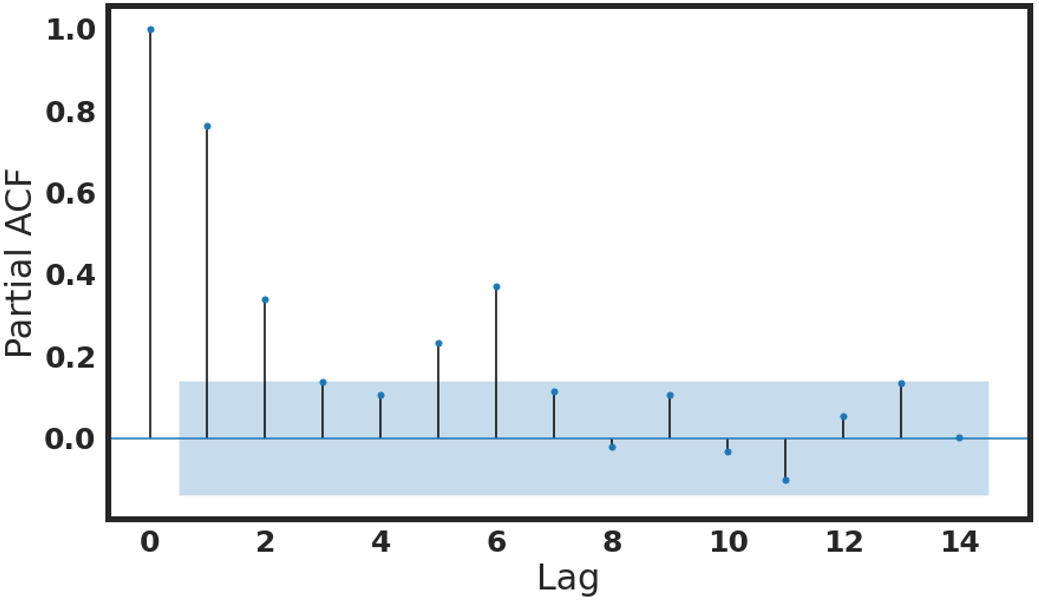}
\includegraphics[width=.24\linewidth]{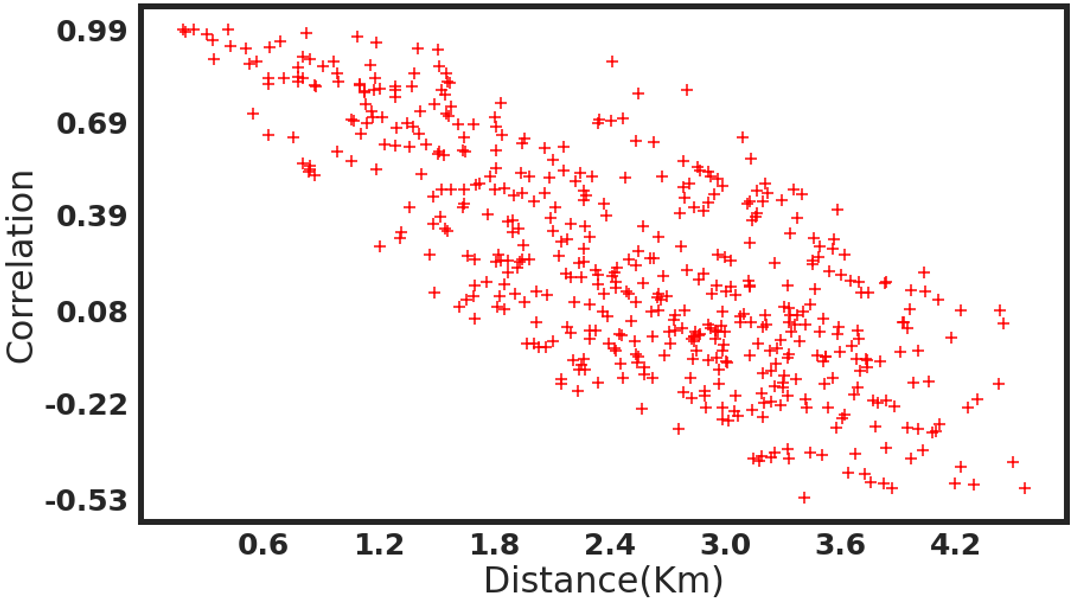}
\caption{\textit{Panel 1, 2, and 3}: PACFs of the time series of confirmed cases in three comunas. The $x$-axis is the time lag in days. The shaded area represents non-significant PACFs. \textit{Panel 4}: Correlation coefficients between the series of confirmed cases density at five arbitrarily chosen locations and those at other 199 locations in their neighborhoods against the separating distances measured in kilometers.}
\label{fig:spatial-temporal-correlation}
\end{figure}

Our preliminary study suggests that the confirmed cases are unevenly distributed across the city and correlated in time and space. In Fig.~\ref{fig:resolutions}, we show the spatial distribution of all the confirmed cases. 
As we can see, most of the reported cases concentrate in the city's center, particularly in Comuna 11. More cases are reported in eastern Cali than in western Cali, which presents a heterogeneous spatial profile of the COVID-19 cases in Cali. The first three panels in Fig.~\ref{fig:spatial-temporal-correlation} show the partial autocorrelation functions (PACF) \citep{PACF} of daily confirmed cases for three comunas in Cali. 
Short lags (less than one week) appear to be highly relevant to the current confirmed cases at each comuna, highlighting a significant temporal dependence.
The last panel of Fig.~\ref{fig:spatial-temporal-correlation} shows the spatial correlation versus the distance between different locations in Cali. Specifically, we investigate the time series of cases occurrence rate (estimated by KDE) at 1,000 arbitrary locations. 
As we can see, a strong spatial correlation is observed in the vicinity of an arbitrary location, while the correlation between two locations weakens with their distance. 

\section{Methodology}

\label{sec:proposed-method}

This section presents our non-stationary spatio-temporal point process model for COVID-19 cases. We first revisit some essential background of spatio-temporal point processes in the following. 
Then we propose a novel point process model with a non-stationary kernel function, which captures complex triggering effects between events in time and space. Lastly, we characterize the influence of city landmarks as an exogenous promotion. 

\subsection{Background: Spatio-temporal point processes}

Spatio-temporal point processes (STPPs) is a popular model for discrete events data that occur in space and time \cite{GONZALEZ2016505,Reinhart2017}. Denote the observation space as $\mathcal{X} = [0, T] \times \mathcal{S} \subseteq \mathbb{R}^+\times \mathbb{R}^2$, where $T$ is the time horizon and $\mathcal{S}$ represents the space of geographic coordinate system (GCS). Each confirmed case is a {\it discrete event} defined by a data tuple $x \coloneqq (t, s)$, where $t \in [0, T]$ is the time when the individual was diagnosed with COVID-19 and $s \in \mathcal{S}$ represents the location of residence of confirmed case.
Let $\mathcal{H}_{t} := \{x_i = (t_i, s_i)|t_i < t\}$ denote the events' history before time $t$. 
Let $\mathbb{N}$ be a counting measure on $[0, T] \times \mathcal{S}$ corresponding to $\mathcal{H}_T$, i.e. for any $S \subset [0, T] \times \mathcal{S}, \mathbb{N}(S) = |S\cap\mathcal{H}_T|$, the number of occurred events in the set $S$. 
For any function $f : [0, T] \times \mathcal{S} \to \mathbb{R}$, the integral w.r.t. the counting measure is defined as
\[
    \int_S f(\tau, r)d\mathbb{N}(\tau, r) = \sum_{(t_i, s_i) \in S \cap \mathcal{H}_T}f(t_i, s_i).
\]
Given the observed history $\mathcal{H}_{t}$, the probability structure of the point process is characterized by the conditional intensity function $\lambda(t, s)$ (for notational simplicity, we omit the dependence on $\mathcal{H}_{t}$), which is defined as 
\begin{equation}
    \lambda(t, s)dt \cdot |B(s, ds)| = \mathbb{E}[d\mathbb{N}(t, s)|\mathcal{H}_{t}].
    \label{eq:original-definition}
\end{equation}
Here $B(s, ds)$ is a ball centered at $s$ in the space $\mathcal{S}$ with radius $ds$, and $|B(s, ds)|$ is the Lebesgue measure. 

Hawkes processes \citep{Self-excitingProcess} is a type of self-exciting point process that captures the triggering effects between events. Assuming that influences from past events are linearly additive to the current event, the conditional intensity for a Hawkes point process takes the form of
\begin{equation}
    \lambda(t, s) = \lambda_0 + \int_0^t\int_{\mathcal{S}}k(t, \tau, s, u)d\mathbb{N}(\tau, u),
    \label{eq:general-form}
\end{equation}
where $\lambda_0 >0$ denotes the background intensity, and $k(t, t^\prime, s, s^\prime)$ is a triggering kernel function that captures the influence of past events on the likelihood of event occurrence at the current time. In this work, we do not assume the kernel function to be positive or shift-invariant (to capture the non-stationary process as we will define later on).

The parameters can be estimated by maximum likelihood estimation (MLE).  Given the observed point pattern $\bm{x}$, we can write the log-likelihood as
\begin{equation}
    \ell(\bm{x}) = \sum_{i=1}^{\mathbb{N}([0, T] \times \mathcal{S})} \log \lambda(t_i, s_i) - \int_{0}^{T}\int_{\mathcal{S}} \lambda(\tau, u)dud\tau,
    \label{eq:log-likelihood}
\end{equation}
where $\mathbb{N}([0, T] \times \mathcal{S})$ is the number of observed events (see the derivation of the log-likelihood in Appendix~\ref{append:derivation-log-likelihood}).

The Epidemic Type Aftershock-Sequences (ETAS) model is one of the most common spatio-temporal point processes \citep{Ogata1988, Ogata1998}, which has been widely adopted in modeling typical spatio-temporal datasets such as earthquakes. ETAS model uses a Gaussian diffusion kernel 
\[
    k(t, t^\prime, s, s^\prime) = \frac{Ce^{-\beta(t - t^\prime)}}{2\pi \sqrt{|\Sigma|}(t - t^\prime)} \cdot \exp{\left \{ -\frac{(s - s^\prime - \mu)^\top \Sigma^{-1} (s - s^\prime - \mu)}{2(t - t ^\prime)} \right \}},
\]
where $\Sigma \equiv \mathrm{diag}(\sigma_x^2, \sigma_y^2)$ is a two-dimensional diagonal matrix representing the covariance of the spatial correlation, $\beta$ is the decaying rate, $\mu$ is the mean shift, and $C$ is a constant.
However, the diffusion kernel is stationary and only depends on the spatio-temporal distance between two events. 
In addition, the kernel assumes the spatial correlation is isotropic and unable to capture complex spatial dependence.  

\subsection{A non-stationary triggering Gaussian kernel}

We introduce a non-stationary triggering kernel, which can vary continuously over space and plays a vital role in modeling the heterogeneous spatial correlation across different regions. For model simplicity and computational efficiency, we adopt the commonly used assumption that the triggering effect of a past event is separable in space and time:
\[
    k(t, t^\prime, s, s^\prime) = \nu(t, t^\prime) \cdot \upsilon(s, s^\prime),
\]
where $\nu(t, t^\prime)$ is a kernel that captures the dependence between time $t$ and $t^\prime$, and $\upsilon(s, s^\prime)$ is a spatial kernel that captures the non-stationary correlation between location $s$ and $s^\prime$. 

\paragraph{A stationary temporal kernel} 

As the virus spreads and affects a significant portion of the population in a short period, we can assume temporal virus transmission is through a shift-invariant kernel with exponential decay:
\[
    \nu(t, t^\prime) =  C e^{-\frac{1}{2\sigma_0^2}(t-t^\prime)^2}, \quad t>t'.
\]
Here $C>0$ is a parameter that controls the magnitude of the kernel, and $\sigma_0 >0$ is a parameter that controls the decaying rate of the event's temporal influences. We assume $t>t'$ to capture the fact that a historical event at time $t'$ impacts the current time $t$ but not vice versa.

\paragraph{A non-stationary spatial kernel} 
The complex nature of the spatial spread of COVID-19 requires a non-homogeneous and non-stationary spatial kernel function in the point process. Given two arbitrary locations $s, s^\prime \in \mathcal S$, we define the spatial kernel $\upsilon(s, s^\prime)$ as a inner product between two feature mappings $\phi_s$ and $\phi_{s'}$, i.e,
\[
    \upsilon(s, s^\prime) = \left <\phi_s, \phi_{s^\prime} \right >, \quad s, s' \in \mathcal S,
\]
where the inner product for functions $\left <f, g \right > \coloneqq \int_{\mathbb{R}^2}f(u)g(u)du$. 
We represent the feature mapping
$\phi_s$ as a weighted sum of a set of $R$ independent kernel-induced feature functions $\{\kappa^{(r)}_s \coloneqq \kappa^{(r)}(s, \cdot)\}_{r=1}^{R}$:
\[
    \phi_s = \sum_{r=1}^{R}w_s^{(r)}\kappa^{(r)}_s,
\]
where 
$\kappa^{(r)}: \mathcal{S} \times \mathcal{S} \rightarrow \mathbb{R}_+$ is a general kernel and
$w_s^{(r)}$ is the corresponding weight of that feature function at location $s$. The location-dependent weight satisfies $\sum_{r=1}^{R}w_s^{(r)} = 1$ at any arbitrary location $s$. 
Hence the spatial kernel can be re-written as
\[
    \upsilon(s, s^\prime) = \sum_{1 \leq r_1, r_2 \leq R} w_s^{(r_1)}w_{s^\prime}^{(r_2)} \left < \kappa_s^{(r_1)}, \kappa_{s^\prime}^{(r_2)} \right >.
\]

\begin{figure}[t]
\centering 
\includegraphics[width=.7\linewidth]{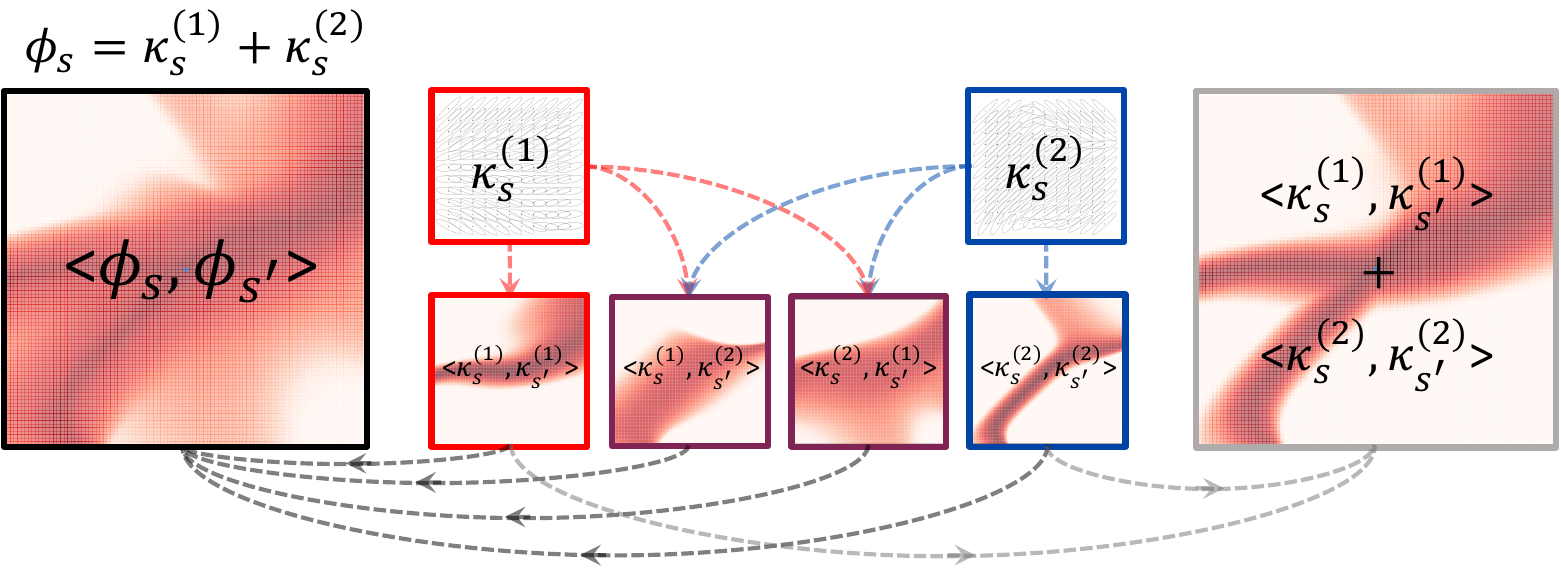}
\caption{
An example of the non-stationary spatial kernel with two feature functions evaluating at location $s$ (the center of the box), i.e., $\upsilon(s, s') = \left< \phi_s, \phi_{s'}\right>,~\forall s' \in \mathcal{S}$, where $\phi_s = \kappa_s^{(1)} + \kappa_s^{(2)}$. 
Two purple boxes in the middle indicate the cross-correlated terms ($\kappa_s^{(1)} \cdot \kappa_{s'}^{(2)}$ and $\kappa_s^{(2)} \cdot \kappa_{s'}^{(1)}$); the red and blue boxes indicate the self-correlated terms ($\kappa_s^{(1)} \cdot \kappa_{s'}^{(1)}$ and $\kappa_s^{(2)} \cdot \kappa_{s'}^{(2)}$).
}
\label{fig:spatial-kernel-illustration}
\end{figure}
The rationale of this design is two-fold: 
(a) Using a linear combination of the product of feature functions enhances the representative power of the spatial kernel. Note that when $r_1 = r_2$, the kernel captures self-correlation (self-similarity of feature functions) and otherwise captures the cross-correlation (similarity between two feature functions). (b) The spatial kernel can also be highly interpretable if $\kappa^{(r)}$ takes a specific parametric form; following the idea in \cite{Higdon1998NonStationarySM, zhu2021early}, we choose $\kappa_s$ to be a Gaussian function centered at $s$ with covariance matrix $\Sigma_s$ since the spatial correlation between two events decays as their distance increases in general. The spatial kernel is specified to be:
\begin{equation}
    \upsilon(s, s^\prime) = \sum_{1\le r_1, r_2 \le R} \frac{w_s^{(r_1)}w_{s^\prime}^{(r_2)}}{2\pi | \Sigma_s^{(r_1)} + \Sigma_{s^\prime}^{(r_2)} |^{\frac{1}{2}}} \exp\left \{-\frac{1}{2}(s - s^\prime)^\top(\Sigma_s^{(r_1)} + \Sigma_{s^\prime}^{(r_2)})^{-1}(s - s^\prime)\right \}.
    \label{eq:spatial-kernel-expression-one-component}
\end{equation}
See detailed derivation of (\ref{eq:spatial-kernel-expression-one-component}) in Appendix~\ref{append:spatial-kernel-derivation}.
Fig.~\ref{fig:spatial-kernel-illustration} gives an example of the spatial kernel with two feature functions.

Now we specify the kernel-induced feature function $\kappa_s$.
According to \cite{Higdon1998NonStationarySM}, there exists a one-to-one mapping between a bivariate normal distribution specified by $\Sigma_s$ and its one standard deviation ellipse. 
Note that $\kappa_s$ is centered at $s$, so the ellipse's center is fixed at $s$. 
Thus we can specify the ellipse by a pair of focus points and the fixed area $A$. The focus points are denoted by $\bm{\psi}_s = (\bm{\psi}_x(s), \bm{\psi}_y(s))$ and $-\bm{\psi}_s = (-\bm{\psi}_x(s), -\bm{\psi}_y(s))$, where $\bm{\psi}_s \in \Psi \subset \mathbb{R}^2$. 
Hence, given $\bm{\psi}_s$ and $A$, the corresponding $\Sigma_s$ can be written as
\[
    \Sigma_s = \tau_z^2 \begin{pmatrix} Q + \frac{\|\bm{\psi}_s\|^2}{2}\cos2\alpha & \frac{\|\bm{\psi}_s\|^2}{2}\sin2\alpha \\ \frac{\|\bm{\psi}_s\|^2}{2}\sin2\alpha & Q - \frac{\|\bm{\psi}_s\|^2}{2}\cos2\alpha \end{pmatrix},
\]
where $Q = \sqrt{4A^2 + \|\bm{\psi}_s\|^4\pi^2}/2\pi, \alpha = \tan^{-1}(\bm{\psi}_y(s) / \bm{\psi}_x(s))$, $\tau_z > 0$ is a scaling parameter
that controls the overall level of the covariance (see the derivation in Appendix~\ref{append:covariance-derivation}). We consider $A$ as a hyper-parameter.

\begin{figure}[t]
\centering 
\includegraphics[width=.7\linewidth]{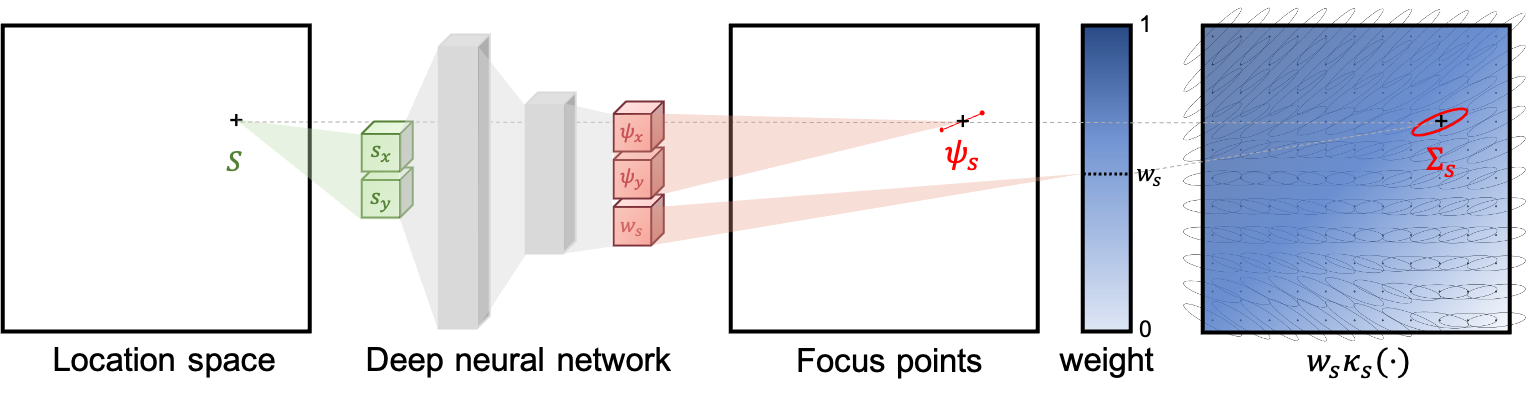}
\caption{An illustration of a deep neural network that maps an arbitrary location $s$ to a spatial kernel, consisting of a feature function $\kappa_s$ (represented through focus points) and weight $w_s$.}
\label{fig:neural-network-illustration}
\end{figure}

\paragraph{Neural network-based kernel representation}

We develop a neural network-based representation for the kernel-induced feature function similar to the idea in \cite{zhu2021early, zhu2022neural}. A key feature of our non-stationary spatial kernel is that for any location $s \in \mathcal{S}$, we can estimate a mapping that obtains the focus point $\bm{\psi}_s$ and the corresponding location-dependent weight $w_s$. To this end, we represent the mapping $\varphi: \mathcal{S} \rightarrow \Psi \times [0, 1]$ from the location to the space of focus points $\Psi$ and the weights $[0, 1]$ using a fully-connected multi-layer neural network. The input of the neural network is the two-dimensional location vector $s$, and the output is the concatenation of the corresponding focus point $\bm{\psi}_s$ and its weight $w_s$. Here, each hidden layer is equipped with a softplus activating function $f(x) = \log(1+e^x)$ (see the detailed specification of the neural network in Section~\ref{sec:results}). Neural networks allow a flexible representation of the covariance and the corresponding kernel-induced feature function due to their well-known universal approximation power. In our implementation, we adopt the same network architecture for all $R$ kernel-induced feature functions, as illustrated in Fig.~\ref{fig:neural-network-illustration}.


\subsection{Exogenous promotion of city landmarks}

To incorporate the influence of city landmarks, we assume each landmark has a constant exogenous promotion to the virus spread at their locations. To achieve this, we adopt an idea similar to \cite{zhu2021spatio} and introduce an additional term to the conditional intensity function $\lambda(t, s)$ \eqref{eq:general-form}: 
\begin{equation}
    \lambda(t, s) = \lambda_0 + 
    \mystrut{4.ex} 
    \sum_{l=1}^{L} \gamma_l g(s | s_l, \Sigma_l) + 
    \mystrut{4.ex} 
    \sum_{(t^\prime, s^\prime) \in \mathcal{H}_t}k(t, t^\prime, s, s^\prime)\ .
    \label{eq:conditional-intensity}
\end{equation}
The second and third terms represent the exogenous promotion at location $s$ and the endogenous excitation at location $s$ and time $t$, respectively. We use $L$ to denote the number of landmarks, and $\gamma_l$ indicates the significance of landmark $l$. We assume that the exogenous effect induced by landmarks decays with distance to them. Hence, the influence of landmark $l$ located at $s_l$ is modeled by a Gaussian function $g(s|s_l,\Sigma_l)$ centered at location $s_l \in \mathcal{S}$ with covariance $\Sigma_l$. Here we define $\Sigma_l \coloneqq \sigma_l^2 \mathbf{I}$, where $\mathbf{I}$ is an identity matrix.

\section{Efficient computation of the log-likelihood function}
\label{sec:efficient-learning}

The log-likelihood of the spatio-temporal point process defined in \eqref{eq:log-likelihood} is often intractable due to the double integral term. Numerical integral can also be expensive: if the number of randomly sampled points in a three-dimensional space is $K$ and the total number of events is $N$, the computational complexity is $\mathcal{O}(KN)$ ($K \gg N$) using commonly-used numerical integration techniques. In our case, we can write the integral term as
\begin{equation}
    \begin{aligned}
        \int_{0}^{T}\int_{\mathcal{S}} \lambda(\tau, u)dud\tau =
        &\lambda_0|\mathcal{S}|T + 
        \int_{0}^{T} \sum_{l=1}^{L}\gamma_l \underbrace{\mystrut{4.ex} \int_{\mathcal{S}} g(u | s_l, \Sigma_l)du}_{(i)}d\tau \\ &+
        \int_{0}^{T} \sum_{t_i < \tau} Ce^{-\frac{1}{2\sigma_{0}^{2}}(\tau-t_i)^2}d\tau \cdot
        \underbrace{\mystrut{4.ex} \int_{\mathcal{S}} \upsilon(u, s_i)du}_{(ii)},
    \end{aligned}
    \label{eq:integral-term}
\end{equation}
where $|\mathcal{S}|$ is the Euclidean area of the city, and evaluating $(i), (ii)$ are difficult in general because 
(a) Both $(i)$ and $(ii)$ require the integral over the geographical space of Cali $\mathcal{S}$, which has an irregular shape; (b) In $(ii)$, $\upsilon(u, s_i)$ is location-dependent and parameterized by a neural network. 

We circumvent these two difficulties by simplifying the calculation of the integral without significantly impacting the model's accuracy: (a) We expand the integration region $\mathcal{S}$ to the entire geographical space $\mathbb{R}^2$ and account for the boundary effect error by $\epsilon_1$. Note that the kernel $g$ or $\kappa_s$ are Gaussian concentrated around $s$ and most events are located in the interior of $\mathcal{S}$ when choosing sufficiently large $\mathcal{S}$. As suggested by \cite{Ogata1998}, such boundary effect is usually negligible due to the decreased activity in the region's edges. (b) We assume the distance between two focus points ($\bm\psi_s$ and $-\bm\psi_s$) at an arbitrary location $s$ is bounded by a threshold $2c$ (which can be obtained by rescaling the output of neural networks); a large distance between focus points leads to an overstretched ellipse, which is unrealistic in practice. Therefore, when performing numerical integration, we approximate the kernel-induced feature function $\kappa_s$ by a standard Gaussian function denoted by $\kappa_{s}^{0}$, which corresponds to a standard deviation ellipse centered at $s$ with area $A$. The relative error of the integral approximation is denoted by $\epsilon_2$.
In short, these two assumptions reduce the double integral \eqref{eq:integral-term} to an analytical form that can be evaluated directly without numerical integration. 

\begin{prop}[Approximation of the integral in the likelihood function]
Assume the area of the corresponding ellipse of $\kappa_s$ is $A$ and the distance between its focus points is restricted to be smaller than $2c$, then the integral in \eqref{eq:integral-term} can be approximated by
\begin{equation}
    \int_{0}^{T}\int_{\mathcal{S}} \lambda(\tau, r)drd\tau = (1 + \epsilon_2) \left[\lambda_0|\mathcal{S}|T + T \sum_{l=1}^{L}\gamma_l + \sqrt{2\pi}C\sigma_0 \sum_{i=1}^{\mathbb{N}([0, T] \times \mathcal{S})}\left\{h\left(\frac{T - t_i}{\sigma_0}\right) - \frac{1}{2} \right\} \right ] - \epsilon_1,
    \label{eq:approximated-integral}
\end{equation}
where the function $h$ is the cumulative density function of the standard normal distribution and $\epsilon_1, \epsilon_2$ are the boundary effect error and the relative error of the integral approximation, respectively.
Ignoring the boundary effect error $\epsilon_1$, the relative error $\epsilon_2 \in (-1, +\infty)$ can be bounded by:
\[
|\epsilon_2| < \max \left \{U - 1, 1 - \frac{1}{U} \right \},
\]
where $U = (\sqrt{4A^2 + c^4\pi^2} + c^2 \pi)/2A$.
(see the proof in Appendix~\ref{append:proof-efficient-learning})
\label{prop:approximation-of-integral}
\end{prop}


\begin{figure}[t]
\centering 
\includegraphics[width=.4\linewidth]{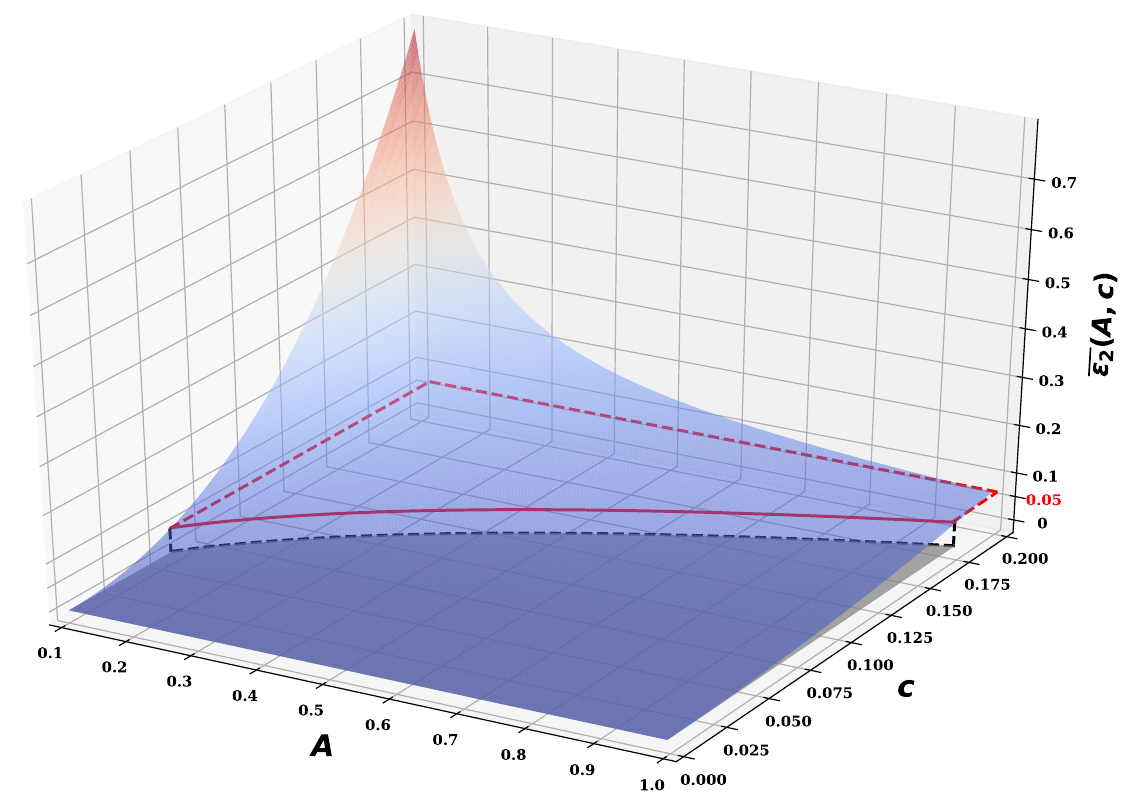}
\caption{Surface plot for the upper bound of the relative error $\overline{\epsilon_2}$ with regards to hyper-parameters $A$ and $c$. The horizontal coordinates represent the value of $A$ and $c$, respectively, and the vertical coordinate represents the value of $\overline{\epsilon_2}(A, c)$. The red solid line is a surface contour valued at $0.05$. 
The grey shaded area in the horizontal plane represents the set of $(A, c)$ that satisfies $\overline{\epsilon_2}(A, c) < 0.05$.
We can observe that the higher the value of $A$ and the lower the value of $c$, the smaller the upper bound of the relative error.}
\label{fig:upper-bound-surface}
\end{figure}
\begin{remark}
Proposition~\ref{prop:approximation-of-integral} leads to a computationally efficient calculation of the integral with complexity $\mathcal{O}(N)$. We denote the upper bound of the relative error as $\overline{\epsilon_2}$ and its dependence on hyper-parameters $A$ and $c$ is illustrated in Fig.~\ref{fig:upper-bound-surface}.
In general, a larger $c$ results in a more expressive spatial kernel but requires a larger $A$ to control the approximation error. In practice, we select $c = 0.1$ and $A = 0.35$ to limit the relative error $\epsilon_2$ under 0.05 and ensure a certain level of expressiveness for the spatial kernel.
\end{remark}

\section{COVID-19 data case study in Cali}
\label{sec:results}


In this section, we present the numerical results for studying the real COVID-19 data in Cali, which is described in Section \ref{sec:data-description}. We first investigate the model's explanatory power by evaluating the in-sample performance and visualizing the estimated kernel-induced feature functions and their corresponding spatial kernel. 
We also study the exogenous effects of the city landmarks. Finally, we compare the out-of-sample predictive performance of the proposed method with four baseline approaches. In this section, $\{\text{MAE} Q_{q}^{in}, \text{MAE} Q_{q}^{out}\}$ denote to the lower $q$-quantile of the mean absolute error (MAE) \citep{Willmott2005MAE} for the in-sample and out-of-sample estimation, respectively.


\begin{figure}[!t]
\centering 
\includegraphics[width=.4\linewidth]{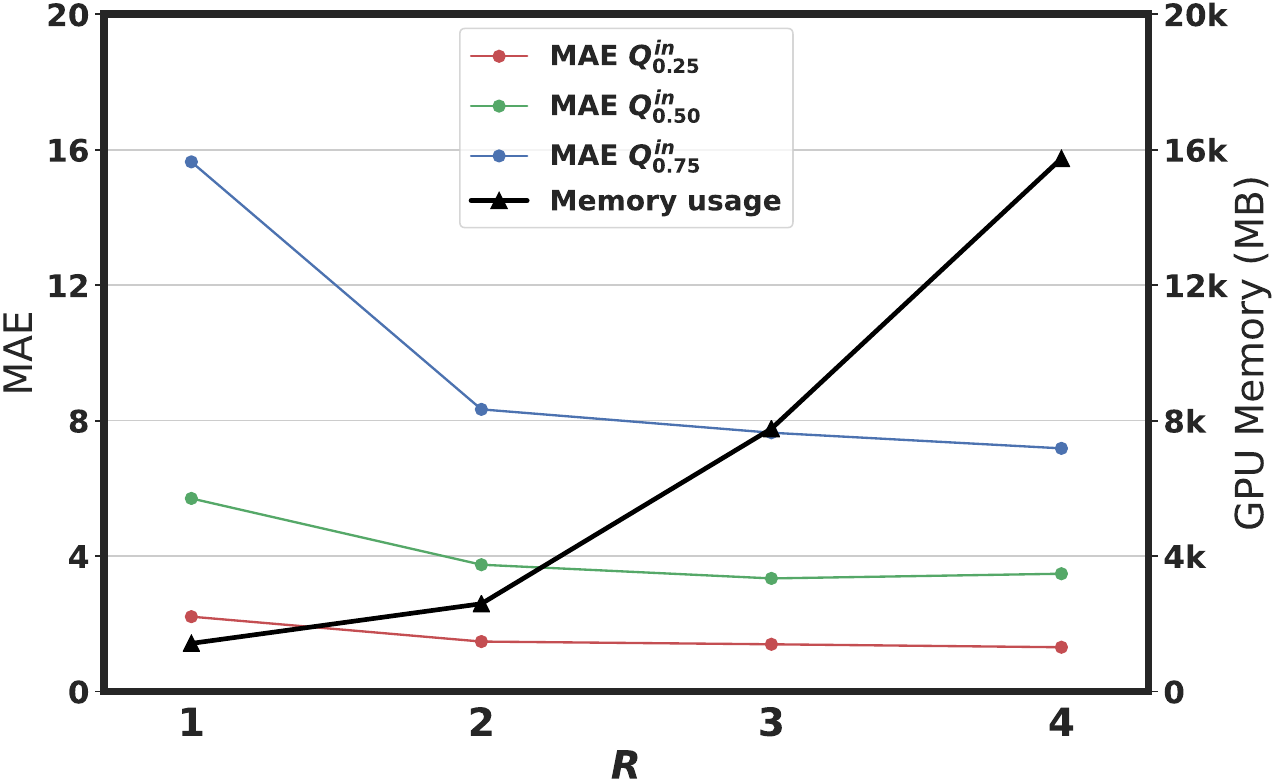}
\includegraphics[width=.55\linewidth]{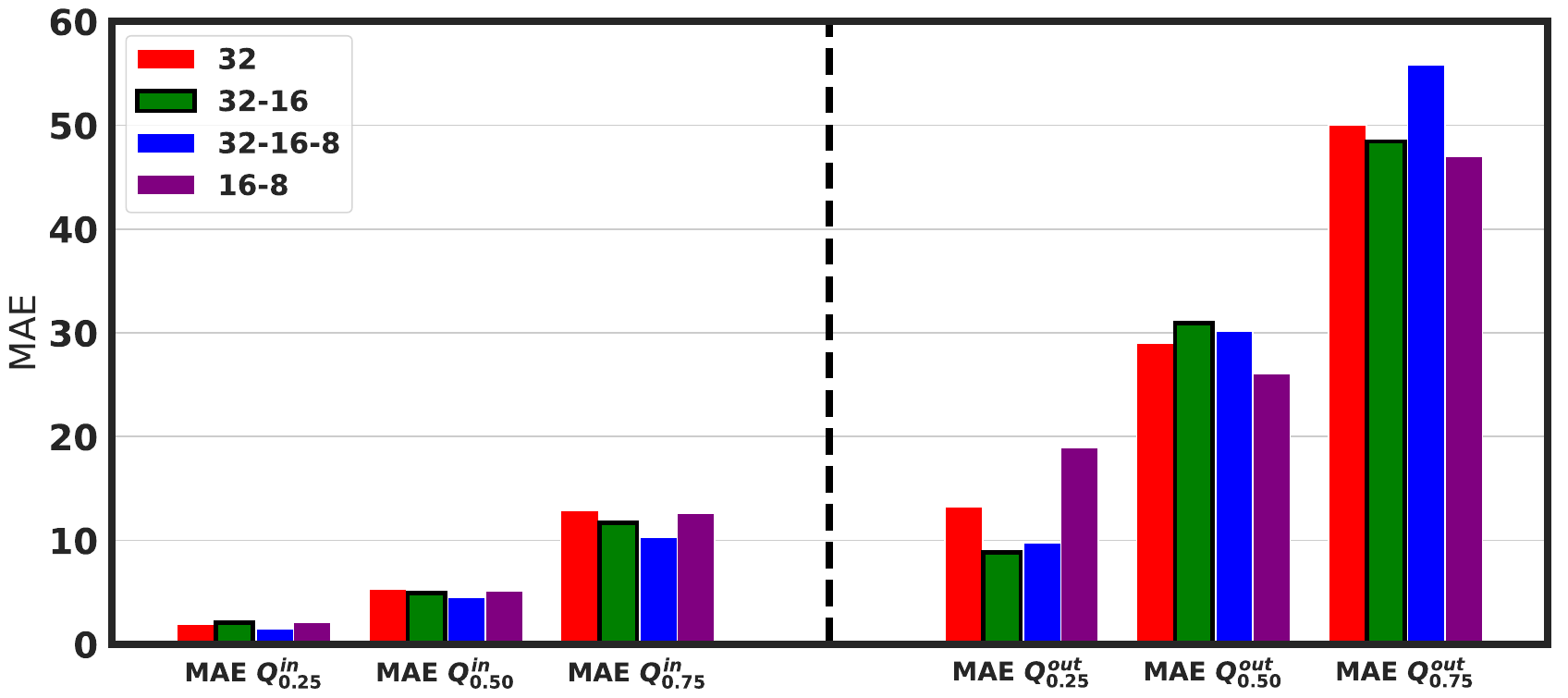}
\caption{
Performance of the proposed model with different numbers of components $R$ in the feature mapping $\phi_s$ or different neural network architectures.
\textit{Left panel}: MAE and GPU memory usage of the in-sample estimations with $R = 1, 2, 3, 4$. 
The red, green, and blue lines represent three different quartiles of MAE for the in-sample estimations, respectively, and the black line represents the increase of GPU memory usage when $R$ grows. 
\textit{Right panel}: MAE of the in-sample and out-of-sample estimations with four different neural network architectures. 
The color code and the corresponding number series represent different neural network structures; 
for example, ``32-16'' indicates a two-hidden-layer neural network, and there are 32 and 16 nodes for each layer. 
The three left groups show the MAEs of in-sample estimation. 
The three right groups show the MAEs of out-of-sample estimation. 
In the following experimental results, we adopt the architecture 32-16.
}
\label{fig:ablation-study}
\end{figure}

Our experimental settings are as follows. We consider a mixture kernel with $R = 3$ components, which achieves the balance between the predictive performance and the computational efficiency according to the results shown in Fig.~\ref{fig:ablation-study}(a). Fig.~\ref{fig:ablation-study}(b) compares the out-of-sample performance for four network architectures; we choose a network architecture that achieves good performance for our data: a two-hidden-layer neural network with 32 and 16 nodes in each hidden layer for each kernel-induced feature function. We select the hyper-parameters $A=0.35$ and $c=0.1$ based on actual needs, and estimate model's parameters $\{\lambda_0, C, \sigma_0, \tau_z, \{\gamma_l\}_{l=1}^{L}, \{\sigma_l\}_{l=1}^{L}, \{\varphi^{(r)}\}_{r=1}^{R}\}$ by solving the maximum likelihood problem via gradient descent 
We train the model with the entire training set in each epoch. The initial learning rate is $1$ and will decay to 0.1 of its last value when there is no likelihood increment for $10$ epochs. The algorithm stops when the likelihood oscillation is less than $1$ for $30$ epochs. We use Adam optimizer \citep{kingma2017adam} for all experiments. In the following, we refer to the proposed framework as a Non-Stationary Spatio-Temporal Point Process (NSSTPP).

\subsection{Model interpretation}

\begin{figure}[!t]
\centering 
\includegraphics[width=.9\textwidth, height=2.5in]{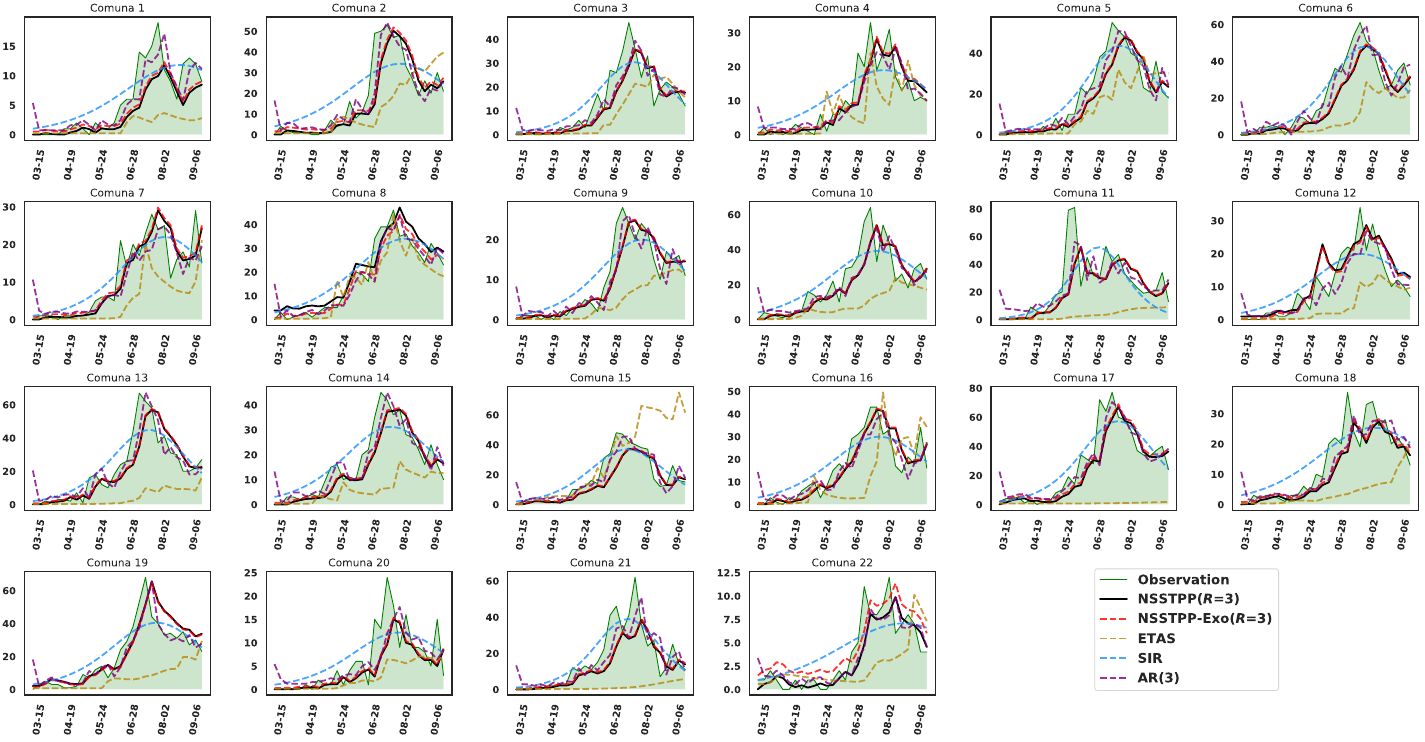}
\caption{Comparison of the proposed model with baseline models. The green lines and shaded areas represent the ground truth. The black and red lines indicate the in-sample estimation of our non-stationary point process model. The yellow, blue, and purple lines represent the in-sample estimation of the ETAS model, SIR model, and AR$(3)$ model, respectively.}
\label{fig:linechart-comparison}
\end{figure}

\begin{table}
\caption{Performance of in-sample estimation. 
Numbers in brackets are standard deviations. The bold values mark the best performance among all models regarding different metrics.
}
\centering
\resizebox{\linewidth}{!}{
\begin{tabular}{lccccc}
\hline
\hline
Models & Log-likelihood($\times 10^4$) & MAE $Q^\text{in}_{0.25}$ & MAE $Q^\text{in}_{0.5}$ & MAE $Q^\text{in}_{0.75}$ \\
\hline
Random & / & $5.000$ & $11.000$ & $18.000$\\
SIR & / & $1.862$ & $3.759$ & $7.391$\\
AR(3) & / & $1.307$ & $2.880$ & $\textbf{6.496}$\\
ETAS & $4.868_{(0.0058)}$ & $1.486_{\color{black}(0.102)}$ & $4.737_{\color{black}(0.240)}$ & $14.895_{\color{black}(0.420)}$ \\
NSSTPP$-$Exo ($R$=1) & $8.671_{(0.0772)}$ & $0.834_{\color{black}(0.087)}$ & $3.145_{\color{black}(0.186)}$ & $7.922_{\color{black}(0.374)}$ \\
NSSTPP$-$Exo ($R$=2) & $9.138_{(0.0886)}$ & $0.806_{\color{black}(0.088)}$ & $2.728_{\color{black}(0.178)}$ & $7.119_{\color{black}(0.361)}$ \\
NSSTPP$-$Exo ($R$=3) & $9.190_{(0.0906)}$ & $0.853_{\color{black}(0.090)}$ & $\textbf{2.613}_{\color{black}(0.165)}$ & $7.000_{\color{black}(0.339)}$ \\
NSSTPP ($R$=3) & $\textbf{9.331}_{(0.0937)}$ & $\textbf{0.797}_{\color{black}(0.085)}$ & $2.620_{\color{black}(0.161)}$ & $6.757_{\color{black}(0.330)}$ \\
\hline  
\hline
\label{tab:in-sample-estimation}
\end{tabular}
}
\end{table}

To evaluate our model's goodness-of-fit, we compare the in-sample estimations of different models on the one-week-ahead number of cases, performed as follows. We first fit the model using the entire 28 weeks of data. The in-sample estimation can then be obtained by feeding the
same data into the fitted model and finding an empirical expectation of the conditional intensity at a given week according to the equation \eqref{eq:conditional-intensity}. 
We compare our model with five baselines that are commonly adopted in modeling infectious epidemics: 
(a) Homogeneous Poisson process (as a sanity check);
(b) Susceptible-Infectious-Recovered (SIR) model; 
(c) Autoregressive (AR) time series model;
(d) Epidemic-type aftershock sequence (ETAS) model;
(e) Our model without exogenous effects (NSSTPP$-$Exo).
See Appendix~\ref{append:additional-experiments} for a detailed review
of the baseline methods and their hyper-parameter choices.
Here we focus on predicting the number of cases for each comuna, which is a subregion
of the city (instead of predicting the occurrence of an event), to compare the predictive accuracy with other
discrete methods. In practice, it can be calculated by $\int_{s\in S^*} \hat\lambda(t^*, s) ds$ using numerical integration, where $t^*$ denotes the time and $S^*$ denotes the region of comuna that we want to predict in.
Fig.~\ref{fig:linechart-comparison} shows the estimated number of cases by different models in each comuna of Cali. 
Table 1 summarizes the in-sample estimation performance measured by two commonly used performance metrics, the log-likelihood and MAE.
In Appendix~\ref{append:additional-experiments}, we first present the results of prediction uncertainty quantification and then compare the performance using different temporal kernels.
The results show that our method outperforms other baseline approaches in both log-likelihood and MAE. Besides, we observe a significant performance gain compared to the ETAS model, which emphasizes the importance of the non-stationarity of the spatial kernel in capturing complex spatio-temporal patterns.

\begin{figure}[!t]
\centering 
\includegraphics[width=.25\linewidth]{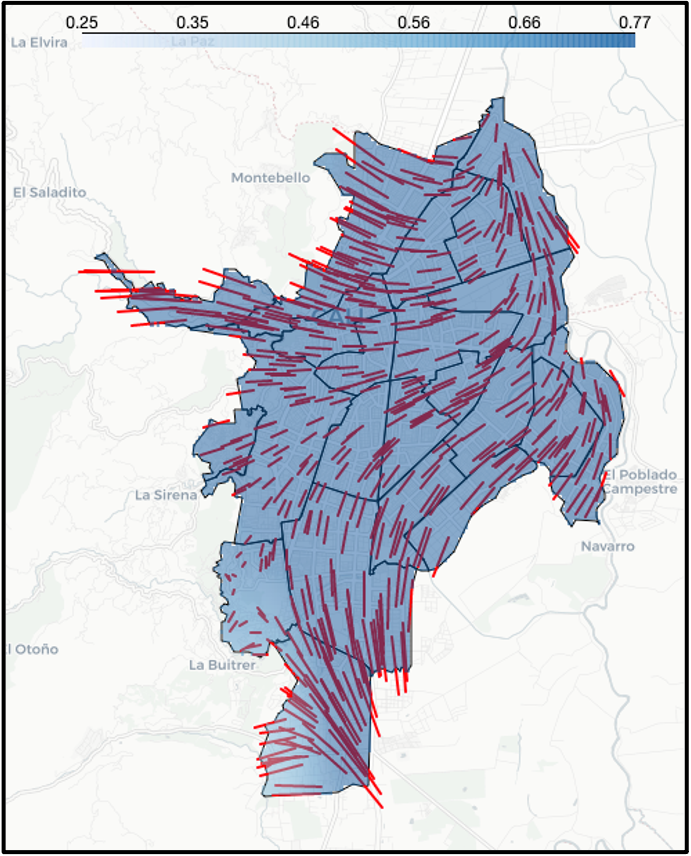}
\includegraphics[width=.25\linewidth]{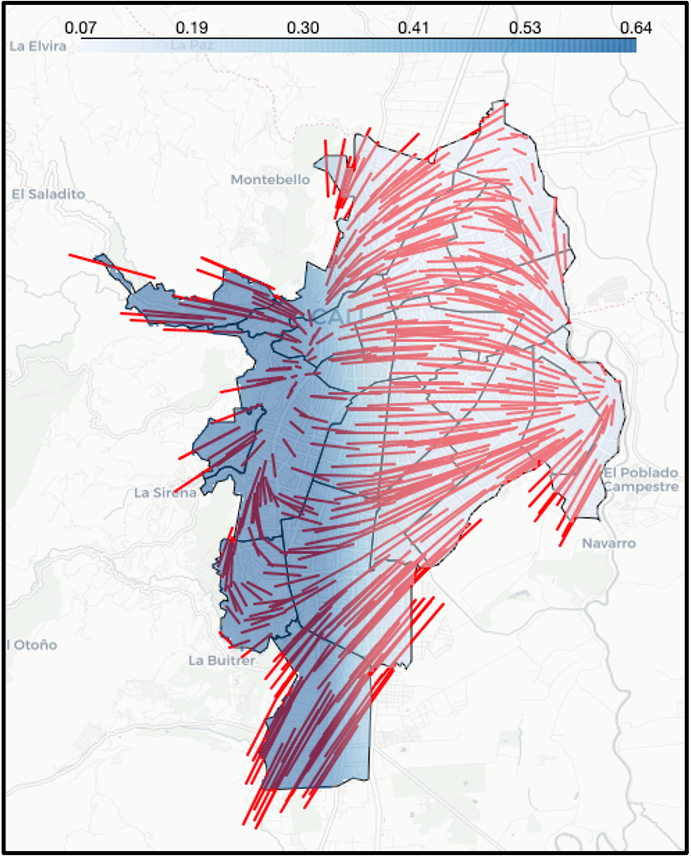}
\includegraphics[width=.25\linewidth]{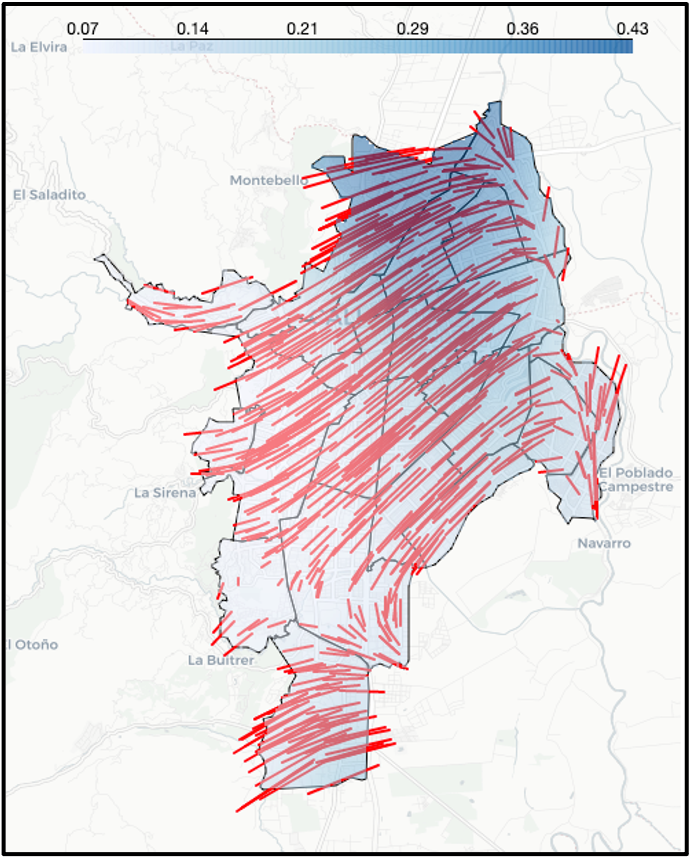}
\caption{Visualization of three learned kernel-induced feature functions over Cali. Each panel shows $\kappa_s^{(1)}, \kappa_s^{(2)}$, and $\kappa_s^{(3)}$ over space, respectively. The line segments plotted above the polygons are edges that connect two focus points of location $s$. The shaded area shows the intensity of weight $w_s^{(r)}$ of each $\kappa_s^{(r)}$ over space. Darker colors mean larger weights.}
\label{fig:kernel-visualization}
\end{figure}

\begin{figure}[!t]
\centering 
\includegraphics[width=.24\linewidth]{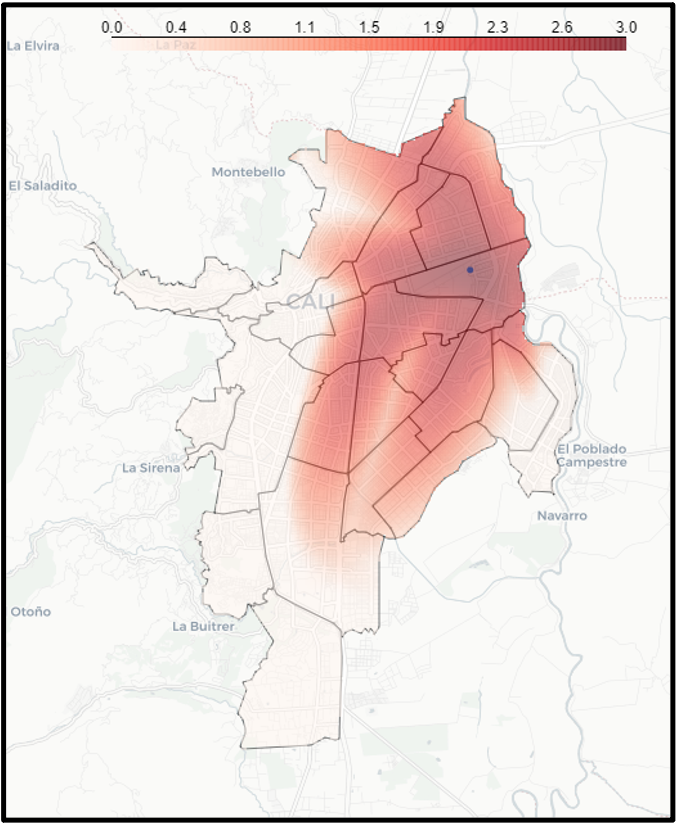}
\includegraphics[width=.24\linewidth]{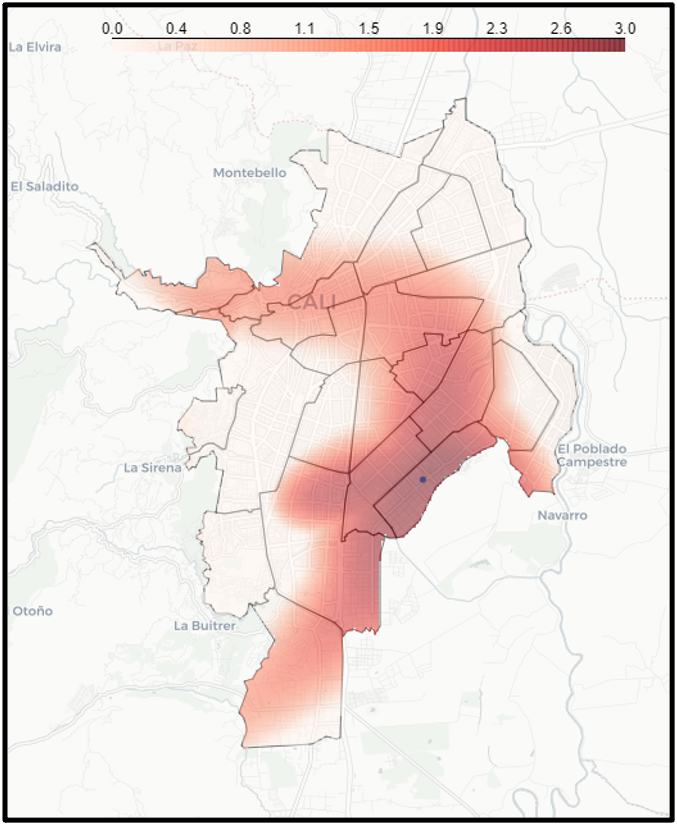}
\includegraphics[width=.24\linewidth]{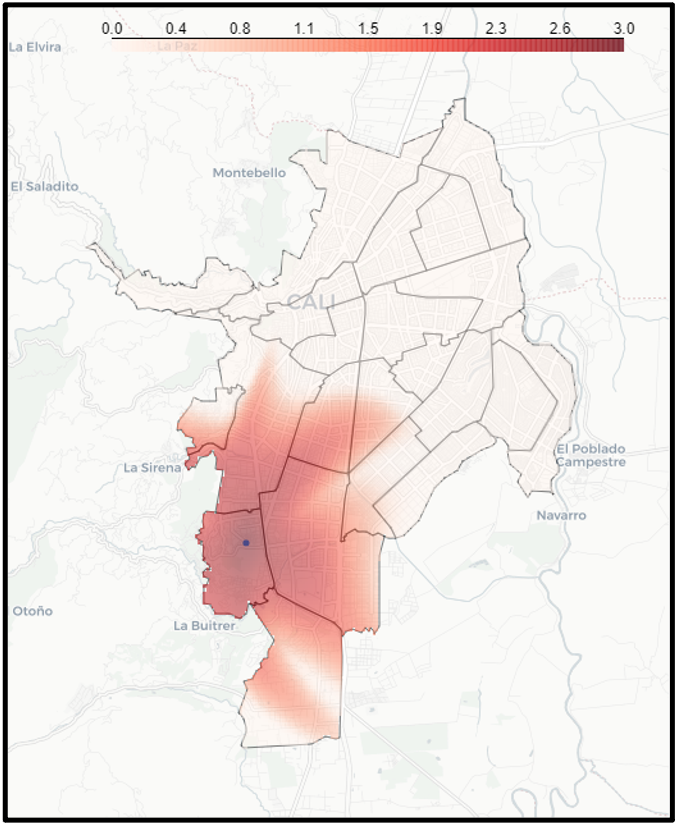}
\includegraphics[width=.24\linewidth]{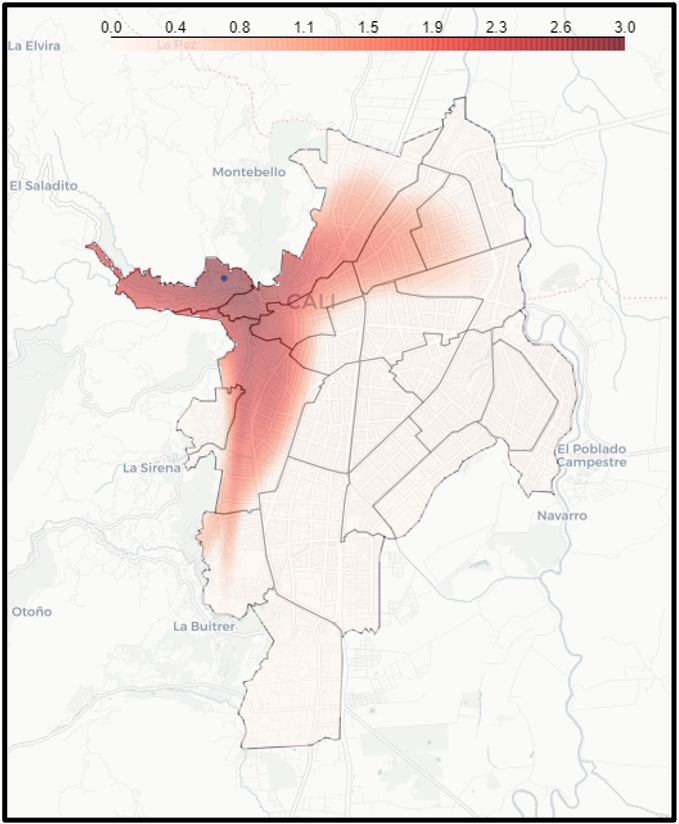}
\caption{Evaluation of the spatial kernel $\upsilon(s, \cdot)$ with $s$ fixed at four typical locations over space: city airport, center of Comuna 15, center of Comuna 18, and center of Comuna 1. These panels intuitively show the spatial influence of the regional hubs located in different parts of the city. The dots represent the fixed location. The color depth indicates the intensity of the kernel value, and the darker color represents a higher kernel value.}
\label{fig:spatial-correlation}
\end{figure}

We study the in-sample explanatory power of our model and interpret the estimation results on the data in Cali. First, we visualize three learned spatial kernel-induced feature functions, which reveal the underlying spatio-temporal transmission dynamics of COVID-19 in Cali, as shown in Fig.~\ref{fig:kernel-visualization}. Recall that at any location $s$,  $\kappa_s^{(r)}$ is a Gaussian kernel with a spatially varying covariance matrix represented by two focus points of its one standard deviation ellipse. Therefore, we connect two focus points of each sampled covariance matrix over space using a red line segment for visualization. The angle and length of each red line can be interpreted as the direction and strength of influence at the particular location. 
The color depth of the background represents the value of the corresponding weight $w_s^{(r)}$ at location $s$ of each $\kappa_s^{(r)}$, indicating the significance of $\kappa_s^{(r)}$ at that location. 
These results suggest that the virus spreads rapidly across the region, following a diagonal direction from the Southwestern to the city's Northeastern. 
We can also observe a more subtle but complicated spreading pattern near the city's border. 

Fig.~\ref{fig:spatial-correlation} visualizes the estimated spatial kernel $\upsilon(s, \cdot)$ given one of its input $s$, which can be treated as the influence of the location $s$. 
Here we present four examples, including the airport, the center of Comuna 1, Comuna 15, and Comuna 18. Each example demonstrates that each location radiates the influence to its surrounding region in a different manner. The results show that the airport significantly influences the other city region as most northern areas have relatively high kernel values. As the most populated community in Cali, Comuna 15 also casts its influences on the city's Southeastern side. In addition, the impact of the location in Comuna 1 extends narrowly to two different directions, which correspond to two major routes in Cali. We note that these examples also emphasize the significance of the non-stationarity of the proposed method.

\begin{figure}[!t]
\centering
\includegraphics[width=.34\linewidth]{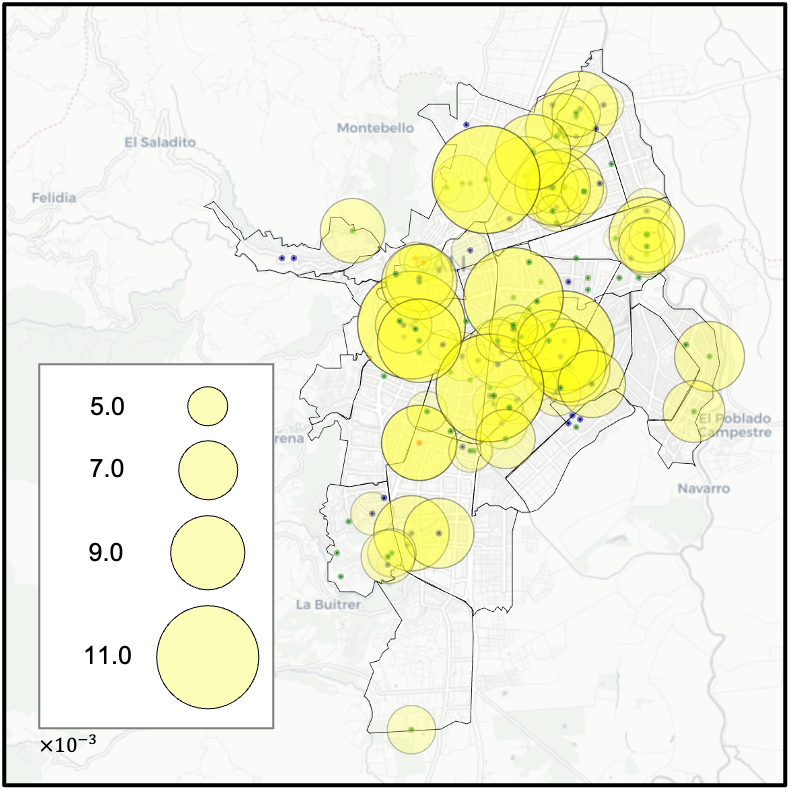}
\includegraphics[width=.34\linewidth]{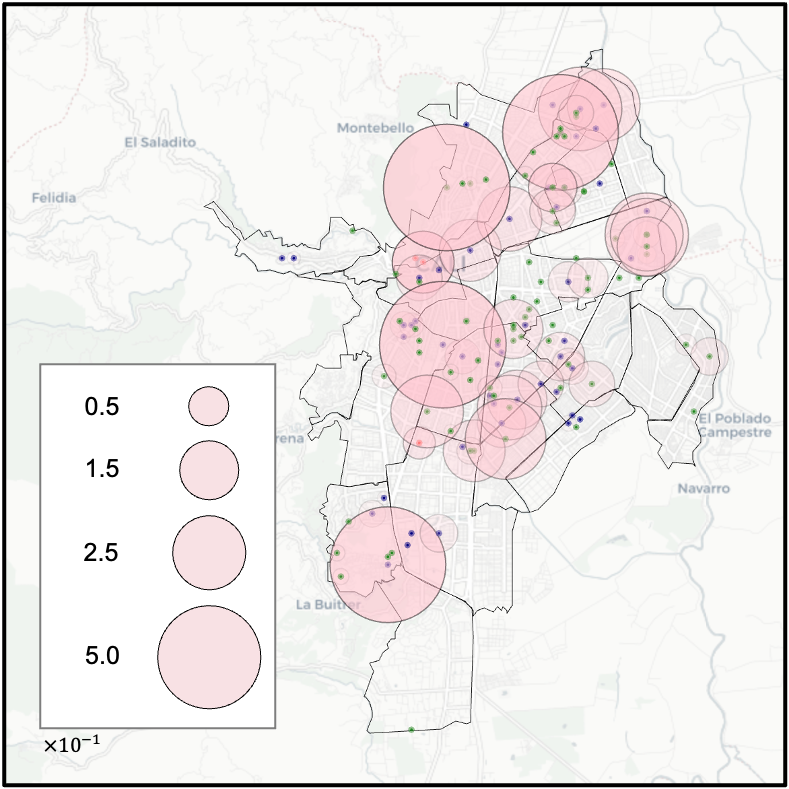}

\includegraphics[width=.34\linewidth]{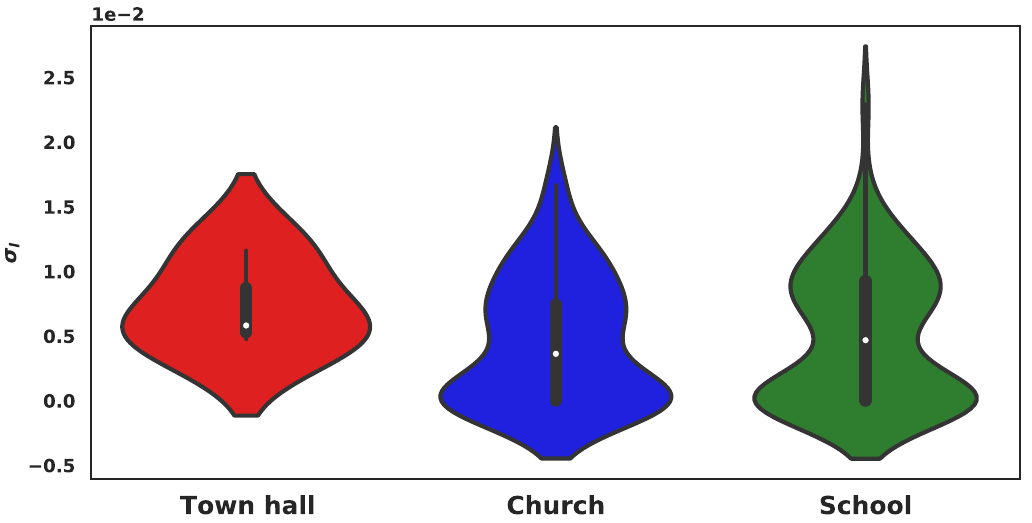}
\includegraphics[width=.34\linewidth]{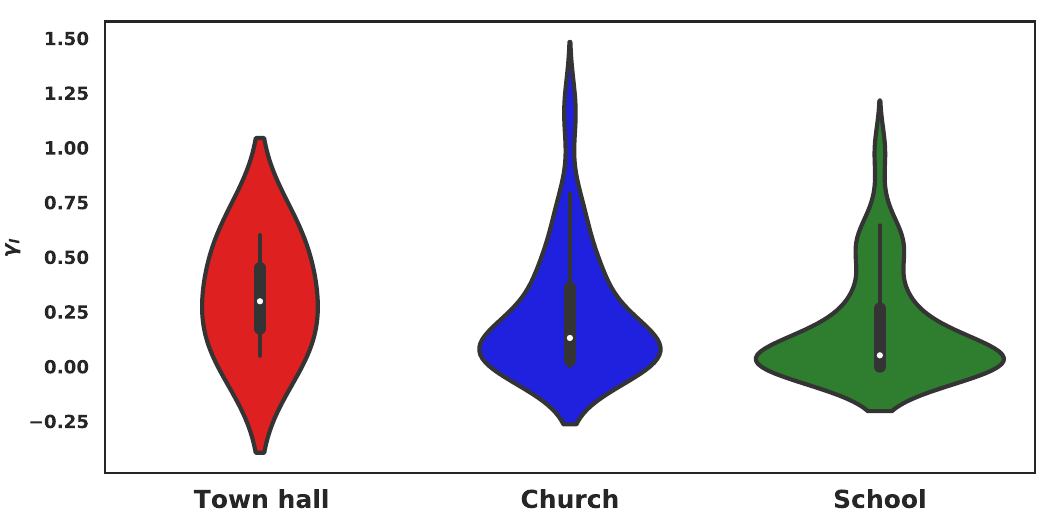}
\caption{
Estimated exogenous effects of landmarks in Cali. 
\textit{Two upper panels} visualize the spatial distribution of learned $\{\sigma_l\}_{l=1}^{L}$ and $\{\gamma_l\}_{l=1}^{L}$, respectively. 
\textit{Two lower panels} show the distributions of learned parameters for different categories of landmarks.  
}
\label{fig:exogenous-effect}
\end{figure}

We also visually and quantitatively examine the exogenous effect of the city landmarks, as shown in Fig.~\ref{fig:exogenous-effect}.
Recall that the exogenous effect of each landmark is assumed to be an isotropic bivariate normal distribution, where $\gamma_l$ and $\sigma_l$ can be interpreted as the \textit{intensity} and the \textit{sphere of influence} of the exogenous effects of landmarks $l$, respectively.
We visualize the learned $\sigma_l$ and $\gamma_l$ on the map of Cali in Fig.~\ref{fig:exogenous-effect}.
We also report the distributions of these two learned parameters for different categories of landmarks. 
As we can observe, the exogenous effects of the landmarks located in the center of the city (the most severely affected areas) tend to have smaller intensities ($\gamma_l$) but larger spheres of influence ($\sigma_l$) on their neighborhood. This result indicates that town halls may have a more significant influence than other landmarks. We see a natural explanation here: The landmarks located at the center receive more people during the day. They act as super-spreaders of the virus, indicated by larger spheres of influence.

%

\subsection{Predictive performance}

\begin{figure}[!t]
\centering 
\includegraphics[width=.24\linewidth]{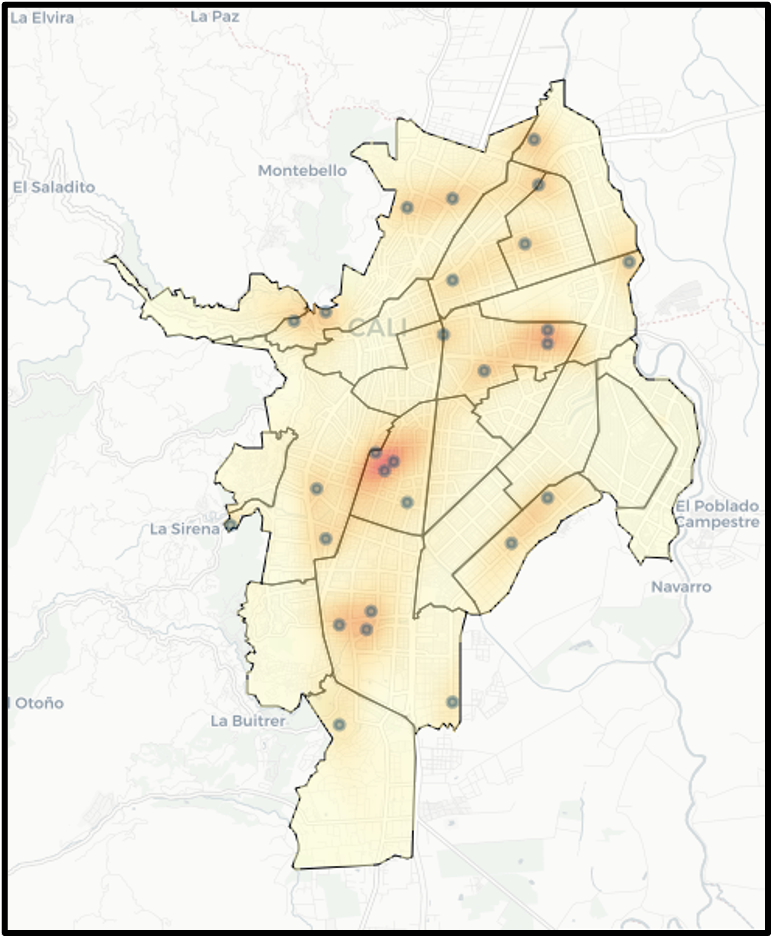}
\includegraphics[width=.24\linewidth]{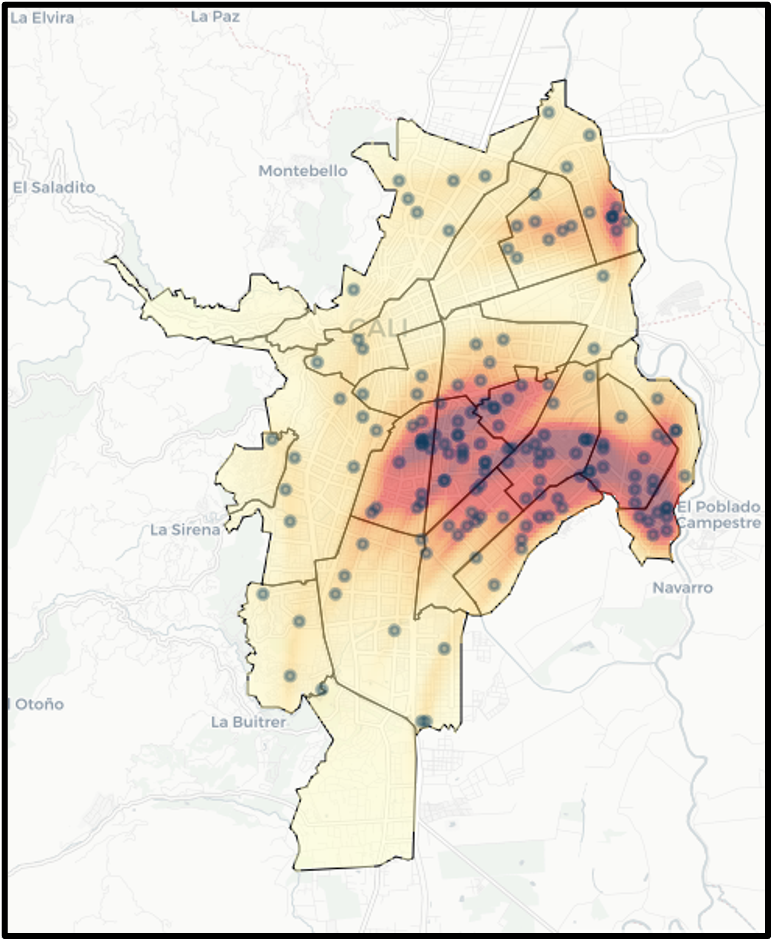}
\includegraphics[width=.24\linewidth]{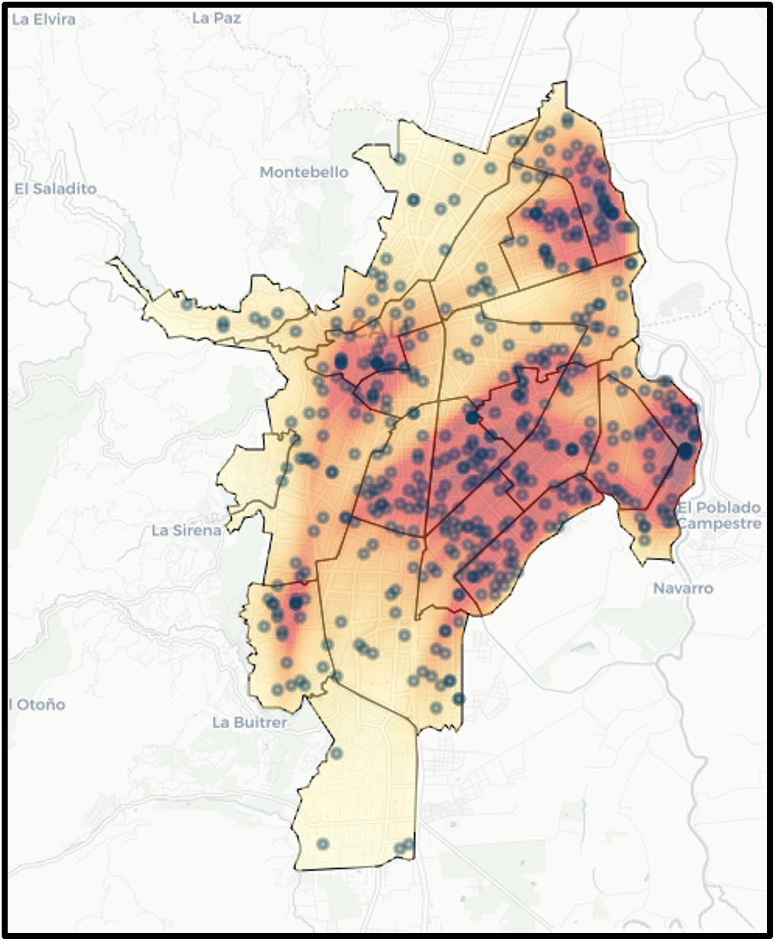}
\includegraphics[width=.24\linewidth]{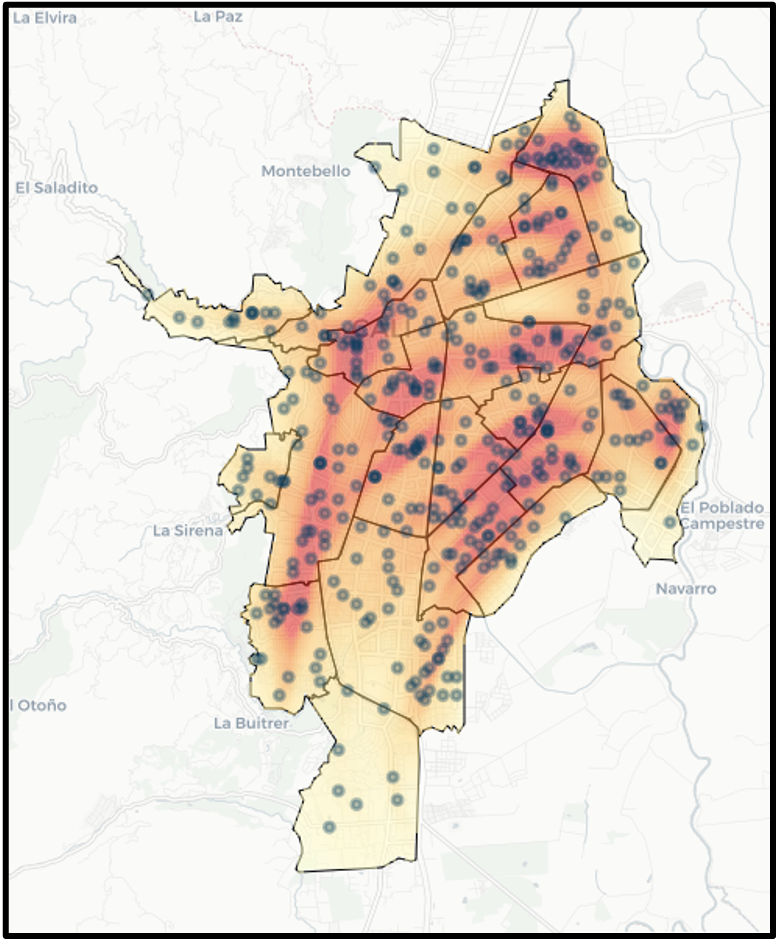}
\caption{
Predicted conditional intensity at four weeks: March 22, May 17, June 28, and August 30. The black dot represents an actual case reported in that week. The color depth indicates the conditional intensity at the corresponding location, and a darker color means a higher risk for citizens to be infected.
}
\label{fig:intensity-map-of-new-cases}
\end{figure}

We assess the model’s predictive power by performing the one-week-ahead out-of-sample prediction of the number of cases. 
The out-of-sample prediction is a one-week-ahead prediction. We randomly hold out the data of one certain week $t^*$ as the testing data and train the model using the data before this week. Then we use the fitted model to predict the conditional intensity function $\hat\lambda(t^*, s)$ for the week $t^*$ over space.
Fig.~\ref{fig:intensity-map-of-new-cases} shows the predicted conditional intensity at four particular weeks, which represent four different stages of the pandemic: (a) the early stage, (b) the week before the first outbreak, (c) the week before the second outbreak, and (d) the week in the stabilized plateau of the pandemic development. 
As we can observe, our method can capture the spatial occurrences of these cases, detect regions with sparsely distributed cases by showing a lower intensity, and show a higher intensity in other regions with densely distributed cases.
To further validate the effectiveness of our model, we implement the out-of-sample prediction for the last four weeks in our data. We report the prediction MAE and corresponding confidence interval for our proposed model (NSSTPP) and other baselines in Table 2. The results validate that our model significantly outperforms other baseline methods for these four weeks, verifying that our model does not overfit data.


\begin{table}
\caption{Performance of out-of-sample estimation for last four weeks. The bold values mark the best performance among all models regarding different metrics.
}
\centering
\resizebox{.8\linewidth}{!}{
\begin{tabular}{lcccc}
\hline
\hline
Models & MAE $Q^\text{out}_{0.25}$ & MAE $Q^\text{out}_{0.5}$ & MAE $Q^\text{out}_{0.75}$ \\
\hline
Random & $10.000$ & $19.000$ & $28.000$\\
SIR & $7.101$ & $12.937$ & $18.809$\\
AR(3) & $3.332$ & $6.571$ & $12.433$\\
ETAS & $4.908_{(0.820)}$ & $12.143_{(0.848)}$ & $22.238_{(0.852)}$ \\
NSSTPP-Exo($R$=1) & $4.783_{(0.661)}$ & $8.944_{(0.748)}$ & $15.834_{(0.843)}$ \\
NSSTPP-Exo($R$=2) & $4.710_{(0.643)}$ & $9.038_{(0.722)}$ & $15.804_{(0.816)}$ \\
NSSTPP-Exo($R$=3) & $4.548_{(0.648)}$ & $8.229_{(0.734)}$ & $15.750_{(0.809)}$ \\
NSSTPP($R$=3) & $\textbf{2.843}_{(0.585)}$ & $\textbf{6.398}_{(0.751)}$ & $\textbf{10.495}_{(0.892)}$ \\
\hline  
\hline
\label{tab:out-of-sample-estimation}
\end{tabular}
}
\end{table}

\section{Conclusions}
\label{sec:conclusion}

Based on an unprecedented fine-grained COVID-19 dataset in Cali, Colombia, we propose a spatio-temporal point process framework equipped with a non-stationary kernel to model epidemic transmission at an individual level. The kernel is composed of a set of kernel-induced feature functions. Each feature function is represented by a neural network aiming to enhance the model flexibility while being interpretable. We also develop an efficient log-likelihood estimation by approximating the double integral using an analytical expression. Our numerical study in Cali has shown that the proposed approach achieves promising predicting performance and the learned model is highly interpretable.

We believe our methodology combines in a natural while novel way theory of point processes with machine learning methods (such as neural networks), providing a unified framework for dealing with highly non-stationary spatio-temporal point patterns. The method is more general than the focused application and can be used, extended, and adapted to several natural phenomena represented by locations in space and time.

There are many ways the proposed method can be extended, and one possibility could be considering non-Gaussian kernels and alternative neural network methods. In any case, the data should always guide these ad-hoc adaptations.

The global results finding in this work for the city of Cali show an increased risk of contracting COVID-19 in the center, northeast, and northeast of the city, which are located in the communes with more unsatisfied basic needs. On the other hand,  in the south of the city, the risk of contagion is lower, and it is an area where people with greater purchasing power live. Considering the locations of the city landmarks in the model as transmission sources is undoubtedly an indispensable tool for predicting the spread of the virus. These outputs are very close to the reality experienced in that region of Colombia during the pandemic. The current official figures for Cali city show significant progress in the fight against COVID-19.





\section*{Acknowledgements}
The authors are grateful to the Municipal Public Health Secretary of Cali, Valle del Cauca, Colombia for providing the COVID-19 data used in this paper.
%
%
The work of Dong, Zhu, and Xie is partially supported by National Science Foundation CCF-1650913, CMMI-2015787, DMS-1938106, DMS-1830210. Francisco J. Rodr\'iguez-Cort\'es has been partially supported by Universidad Nacional de Colombia, HERMES projects, Grant/Award Number: 51279.




\bibliographystyle{apalike}
\bibliography{refs}

\begin{thebibliography}{}

\bibitem[Agosto and Giudici, 2020]{Agosto2020PoissonAR}
Agosto, A. and Giudici, P. (2020).
\newblock A poisson autoregressive model to understand {COVID}-19 contagion
  dynamics.
\newblock {\em Risks}, 8(3):77.

\bibitem[Angulo et~al., 2013]{SpatialtemporalSIR}
Angulo, J., Yu, H.-L., Langousis, A., Kolovos, A., Wang, J., Madrid, A.~E., and
  Christakos, G. (2013).
\newblock Spatiotemporal infectious disease modeling: a {BME}-{SIR} approach.
\newblock {\em PloS one}, 8(9).

\bibitem[Bai et~al., 2020]{Bai2020NonstatSIR}
Bai, Y., Safikhani, A., and Michailidis, G. (2020).
\newblock Non-stationary spatio-temporal modeling of {COVID}-19 progression in
  the u.s.
\newblock {\em medRXiv preprint}.

\bibitem[Berry et~al., 2020]{berry2020open}
Berry, I., Soucy, J.-P.~R., Tuite, A., and Fisman, D. (2020).
\newblock Open access epidemiologic data and an interactive dashboard to
  monitor the covid-19 outbreak in canada.
\newblock {\em Cmaj}, 192(15):E420--E420.

\bibitem[Bizzarri et~al., 2020]{trackCOVID-19Italy}
Bizzarri, M., Di~Traglia, M., Giuliani, A., Vestri, A., Fedeli, V., and
  Prestininzi, A. (2020).
\newblock New statistical ri index allow to better track the dynamics of
  covid-19 outbreak in italy.
\newblock {\em Scientific Reports}, 10.

\bibitem[Brand{\'e}n et~al., 2020]{branden2020residential}
Brand{\'e}n, M., Aradhya, S., Kolk, M., H{\"a}rk{\"o}nen, J., Drefahl, S.,
  Malmberg, B., Rostila, M., Cederstr{\"o}m, A., Andersson, G., and Mussino, E.
  (2020).
\newblock Residential context and covid-19 mortality among adults aged 70 years
  and older in stockholm: a population-based, observational study using
  individual-level data.
\newblock {\em The Lancet Healthy Longevity}, 1(2):e80--e88.

\bibitem[Brockwell and Davis, 1991]{PACF}
Brockwell, P.~J. and Davis, R.~A. (1991).
\newblock Estimation of the mean and the autocovariance function.
\newblock In {\em Springer Series in Statistics}, pages 218--237. Springer New
  York.

\bibitem[Chiang et~al., 2020]{CovariatesHawkesModeling}
Chiang, W.-H., Liu, X., and Mohler, G. (2020).
\newblock Hawkes process modeling of covid-19 with mobility leading indicators
  and spatial covariates.
\newblock {\em medRxiv}.

\bibitem[Chriscaden, 2020]{COVID-19_intro}
Chriscaden, K. (2020).
\newblock Impact of covid-19 on people's livelihoods, their health and our food
  systems.
\newblock
  \url{https://www.who.int/news/item/13-10-2020-impact-of-covid-19-on-people's-livelihoods-their-health-and-our-food-systems}.

\bibitem[Du et~al., 2016]{RMTPP}
Du, N., Dai, H., Trivedi, R., Upadhyay, U., Gomez-Rodriguez, M., and Song, L.
  (2016).
\newblock Recurrent marked temporal point processes.
\newblock In {\em Proceedings of the 22nd {ACM} {SIGKDD} International
  Conference on Knowledge Discovery and Data Mining}. {ACM}.

\bibitem[Farajtabar et~al., 2017]{farajtabar2017fake}
Farajtabar, M., Yang, J., Ye, X., Xu, H., Trivedi, R., Khalil, E., Li, S.,
  Song, L., and Zha, H. (2017).
\newblock Fake news mitigation via point process based intervention.
\newblock In {\em International conference on machine learning}, pages
  1097--1106. PMLR.

\bibitem[Fu et~al., 2020]{fu2020tocilizumab}
Fu, B., Xu, X., and Wei, H. (2020).
\newblock Why tocilizumab could be an effective treatment for severe covid-19?
\newblock {\em Journal of translational medicine}, 18(1):1--5.

\bibitem[Gajardo and M\"{u}ller, 2021]{Gajardo2021PPforCOVID19}
Gajardo, {\'{A}}. and M\"{u}ller, H.-G. (2021).
\newblock Point process models for {COVID}-19 cases and deaths.
\newblock {\em Journal of Applied Statistics}, pages 1--16.

\bibitem[Giudici et~al., 2023]{Giudici2021NetworkPP}
Giudici, P., Pagnottoni, P., and Spelta, A. (2023).
\newblock {Network self-exciting point processes to measure health impacts of
  COVID-19}.
\newblock {\em Journal of the Royal Statistical Society Series A: Statistics in
  Society}.
\newblock qnac006.

\bibitem[Gonz\'alez et~al., 2016]{GONZALEZ2016505}
Gonz\'alez, J.~A., Rodr\'iguez-Cort\'es, F.~J., Cronie, O., and Mateu, J.
  (2016).
\newblock Spatio-temporal point process statistics: A review.
\newblock {\em Spatial Statistics}, 18:505--544.

\bibitem[Guenther et~al., 2020]{superspreadinGermany}
Guenther, T., Czech-Sioli, M., Indenbirken, D., Robitailles, A., Tenhaken, P.,
  Exner, M., Ottinger, M., Fischer, N., Grundhoff, A., and Brinkmann, M.
  (2020).
\newblock Investigation of a superspreading event preceding the largest meat
  processing plant-related sars-coronavirus 2 outbreak in germany.
\newblock {\em SSRN Journal}.

\bibitem[Guo et~al., 2020]{guo2020tocilizumab}
Guo, C., Li, B., Ma, H., Wang, X., Cai, P., Yu, Q., Zhu, L., Jin, L., Jiang,
  C., Fang, J., et~al. (2020).
\newblock Tocilizumab treatment in severe covid-19 patients attenuates the
  inflammatory storm incited by monocyte centric immune interactions revealed
  by single-cell analysis.
\newblock {\em bioRxiv}.

\bibitem[Harko et~al., 2014]{SIR}
Harko, T., Lobo, F. S.~N., and Mak, M.~K. (2014).
\newblock Exact analytical solutions of the susceptible-infected-recovered
  (sir) epidemic model and of the sir model with equal death and birth rates.
\newblock {\em Applied Mathematics and Computation}, 236.

\bibitem[Hawkes, 1971]{Self-excitingProcess}
Hawkes, A.~G. (1971).
\newblock Spectra of some self-exciting and mutually exciting point processes.
\newblock {\em Biometrika}, 58(1):83--90.

\bibitem[Hendry and Pretis, 2016]{hendry2016all}
Hendry, D.~F. and Pretis, F. (2016).
\newblock All change! the implications of non-stationarity for empirical
  modelling, forecasting and policy.
\newblock {\em SSRN Journal}.

\bibitem[Higdon et~al., 2022]{Higdon1998NonStationarySM}
Higdon, D., Swall, J., and Kern, J. (2022).
\newblock Non-stationary spatial modeling.
\newblock {\em arXiv preprint arXiv:2212.08043}.

\bibitem[Hochreiter and Schmidhuber, 1997]{Hochreiter1997LSTM}
Hochreiter, S. and Schmidhuber, J. (1997).
\newblock Long short-term memory.
\newblock {\em Neural Computation}, 9(8):1735--1780.

\bibitem[{Institute for Health Metrics and Evaluation}, 2020]{2020IHME-CDC}
{Institute for Health Metrics and Evaluation} (2020).
\newblock Modeling {COVID}-19 scenarios for the united states.
\newblock {\em Nature Medicine}, 27(1):94--105.

\bibitem[James et~al., 2021]{web2021high}
James, A., Eagle, L., Phillips, C., Hedges, D.~S., Bodenhamer, C., Brown, R.,
  Wheeler, J.~G., and Kirking, H. (2021).
\newblock High covid-19 attack rate among attendees at events at a church —
  arkansas, march 2020.
\newblock \url{https://www.cdc.gov/mmwr/volumes/69/wr/mm6920e2.htm}.

\bibitem[Jasper et~al., 2020]{COVID19transLancet}
Jasper, F.-W.~C., Shuofeng, Y., Kin-Hang, K., Kelvin, K.-W.~T., Hin, C., Jin,
  Y., and et~al. (2020).
\newblock A familial cluster of pneumonia associated with the 2019 novel
  coronavirus indicating person-to-person transmission: a study of a family
  cluster.
\newblock {\em Lancet}.

\bibitem[{John Hopkins University}, 2020]{COVID19JHU}
{John Hopkins University} (2020).
\newblock Coronavirus resource center.
\newblock \url{https://coronavirus.jhu.edu}.

\bibitem[Kingma and Ba, 2014]{kingma2017adam}
Kingma, D.~P. and Ba, J. (2014).
\newblock Adam: A method for stochastic optimization.
\newblock {\em arXiv preprint arXiv:1412.6980}.

\bibitem[Korolev, 2021]{SEIRDinCOVID19}
Korolev, I. (2021).
\newblock Identification and estimation of the seird epidemic model for
  covid-19.
\newblock {\em Journal of econometrics}, 220(1).

\bibitem[Kraemer, 2020]{IntegerModel}
Kraemer, M. U. G. e.~a. (2020).
\newblock The effect of human mobility and control measures on the covid-19
  epidemic in china.
\newblock {\em Science}.

\bibitem[Lang et~al., 2007]{lang2007adaptive}
Lang, T., Plagemann, C., and Burgard, W. (2007).
\newblock Adaptive non-stationary kernel regression for terrain modeling.
\newblock In {\em Robotics: Science and Systems}, volume~6. Citeseer.

\bibitem[Leclerc et~al., 2020]{indoorinfection}
Leclerc, Q.~J., Fuller, N.~M., Knight, L.~E., Funk, S., and Knight, G.~M.
  (2020).
\newblock What settings have been linked to sars-cov-2 transmission clusters?
\newblock {\em Wellcome Open Research}.

\bibitem[Li et~al., 2021]{ReinformentSTPP}
Li, S., Wang, L., Chen, X., Fang, Y., and Song, Y. (2021).
\newblock Understanding the spread of {COVID-19} epidemic: {A} spatio-temporal
  point process view.
\newblock {\em CoRR}, abs/2106.13097.

\bibitem[Lin et~al., 2020]{ConceptualModel}
Lin, Q., Zhao, S., Gao, D., Lou, Y., Yang, S., Musa, S.~S., Wang, M.~H., Cai,
  Y., Wang, W., Yang, L., and He, D. (2020).
\newblock A conceptual model for the coronavirus disease 2019 (covid-19)
  outbreak in wuhan, china with individual reaction and governmental action.
\newblock {\em International Journal of Infectious Diseases}.

\bibitem[Loli~Piccolomini and Zama, 2020]{fSEIRDinCOVID19}
Loli~Piccolomini, E. and Zama, F. (2020).
\newblock Monitoring italian covid-19 spread by a forced seird model.
\newblock {\em PloS one}, 15(8).

\bibitem[L{\'o}pez-Feldman et~al., 2021]{lopez2021air}
L{\'o}pez-Feldman, A., Heres, D., and Marquez-Padilla, F. (2021).
\newblock Air pollution exposure and covid-19: a look at mortality in mexico
  city using individual-level data.
\newblock {\em Science of the Total Environment}, 756:143929.

\bibitem[Mamode~Khan et~al., 2020]{ARwithCovariates}
Mamode~Khan, N., Soobhug, A.~D., and Heenaye-Mamode~Khan, M. (2020).
\newblock Studying the trend of the novel coronavirus series in mauritius and
  its implications.
\newblock {\em PloS one}, 15(7).

\bibitem[Mei and Eisner, 2016]{NeuralHawkes}
Mei, H. and Eisner, J. (2016).
\newblock The neural hawkes process: {A} neurally self-modulating multivariate
  point process.
\newblock {\em CoRR}, abs/1612.09328.

\bibitem[Nande et~al., 2020]{DynSocialDistancing}
Nande, A., Adlam, B., Sheen, J., Levy, M.~Z., , and Hill, A.~L. (2020).
\newblock Dynamics of covid-19 under social distancing measures are driven by
  transmission network structure.
\newblock {\em medRxiv}.

\bibitem[{New York Times}, 2020]{COVID19NYTimes}
{New York Times} (2020).
\newblock Coronavirus data in the united states.
\newblock
  \url{https://www.nytimes.com/article/coronavirus-county-data-us.html}.

\bibitem[{Northeastern University, Laboratory for the Modeling of Biological
  and Socio-technical Systems}, 2021]{MOBS}
{Northeastern University, Laboratory for the Modeling of Biological and
  Socio-technical Systems} (2021).
\newblock {COVID}-19 modeling.

\bibitem[Ogata, 1988]{Ogata1988}
Ogata, Y. (1988).
\newblock Statistical models for earthquake occurrences and residual analysis
  for point processes.
\newblock {\em Journal of the American Statistical Association}, 83(401):9--27.

\bibitem[Ogata, 1998]{Ogata1998}
Ogata, Y. (1998).
\newblock Space-time point-process models for earthquake occurrences.
\newblock {\em Annals of the Institute of Statistical Mathematics},
  50(2):379--402.

\bibitem[{Presidency of the Republic of Colombia.}, 2020]{COVID19COL}
{Presidency of the Republic of Colombia.} (2020).
\newblock Decrees during the {COVID}-19 pandemic.
\newblock \url{https://coronaviruscolombia.gov.co/Covid19/decretos.html}.

\bibitem[Reinhart, 2018]{Reinhart2017}
Reinhart, A. (2018).
\newblock A review of self-exciting spatio-temporal point processes and their
  applications.
\newblock {\em Statistical Science}, 33(3):299--318.

\bibitem[Remes et~al., 2017]{Remes2017NonStationarySK}
Remes, S., Heinonen, M., and Kaski, S. (2017).
\newblock Non-stationary spectral kernels.
\newblock {\em Advances in neural information processing systems}, 30.

\bibitem[Rizoiu et~al., 2017]{Rizoiu_Exogenous_Media}
Rizoiu, M.-A., Xie, L., Sanner, S., Cebrian, M., Yu, H., and Van~Hentenryck, P.
  (2017).
\newblock Expecting to be hip: Hawkes intensity processes for social media
  popularity.
\newblock {\em Proceedings of the 26th International Conference on World Wide
  Web}.

\bibitem[Sahai and Khurshid, 1993]{Sahai1993UQ}
Sahai, H. and Khurshid, A. (1993).
\newblock Confidence intervals for the mean of a poisson distribution: A
  review.
\newblock {\em Biometrical Journal}, 35(7):857--867.

\bibitem[Triaccaa and Triacca, 2021]{ARFirst-order}
Triaccaa, M. and Triacca, U. (2021).
\newblock Forecasting the number of confirmed new cases of covid-19 in italy
  for the period from 19 may to 2 june 2020.
\newblock {\em Infectious Disease Modelling}.

\bibitem[Vasudevan et~al., 2011]{vasudevan2011non}
Vasudevan, S., Ramos, F., Nettleton, E., and Durrant-Whyte, H. (2011).
\newblock Non-stationary dependent gaussian processes for data fusion in
  large-scale terrain modeling.
\newblock In {\em 2011 IEEE International Conference on Robotics and
  Automation}, pages 1875--1882. IEEE.

\bibitem[Vaswani et~al., 2017]{vaswani2017attention}
Vaswani, A., Shazeer, N., Parmar, N., Uszkoreit, J., Jones, L., Gomez, A.~N.,
  Kaiser, {\L}., and Polosukhin, I. (2017).
\newblock Attention is all you need.
\newblock In {\em Advances in neural information processing systems}, pages
  5998--6008.

\bibitem[Wikipedia, 2021]{Cali_intro}
Wikipedia (2021).
\newblock Cali.
\newblock \url{https://en.wikipedia.org/wiki/Cali}.

\bibitem[Willmott and Matsuura, 2005]{Willmott2005MAE}
Willmott, C. and Matsuura, K. (2005).
\newblock Advantages of the mean absolute error ({MAE}) over the root mean
  square error ({RMSE}) in assessing average model performance.
\newblock {\em Climate Research}, 30:79--82.

\bibitem[Woody et~al., 2020]{woody2020projections}
Woody, S., Tec, M.~G., Dahan, M., Gaither, K., Lachmann, M., Fox, S., Meyers,
  L.~A., and Scott, J.~G. (2020).
\newblock Projections for first-wave covid-19 deaths across the us using
  social-distancing measures derived from mobile phones.
\newblock {\em medRxiv}.

\bibitem[Zhang et~al., 2020]{AttentiveHawkes}
Zhang, Q., Lipani, A., Kirnap, O., and Yilmaz, E. (2020).
\newblock Self-attentive {H}awkes process.
\newblock In III, H.~D. and Singh, A., editors, {\em Proceedings of the 37th
  International Conference on Machine Learning}, volume 119 of {\em Proceedings
  of Machine Learning Research}, pages 11183--11193. PMLR.

\bibitem[Zhu et~al., 2021a]{zhu2021highresolution}
Zhu, S., Bukharin, A., Xie, L., Santillana, M., Yang, S., and Xie, Y. (2021a).
\newblock High-resolution spatio-temporal model for county-level covid-19
  activity in the us.
\newblock {\em ACM Transactions on Management Information Systems (TMIS)},
  12(4):1--20.

\bibitem[Zhu et~al., 2022a]{zhu2021early}
Zhu, S., Bukharin, A., Xie, L., Yamin, K., Yang, S., Keskinocak, P., and Xie,
  Y. (2022a).
\newblock Early detection of covid-19 hotspots using spatio-temporal data.
\newblock {\em IEEE Journal of Selected Topics in Signal Processing},
  16(2):250--260.

\bibitem[Zhu et~al., 2021b]{zhu2021spatio}
Zhu, S., Ding, R., Zhang, M., Van~Hentenryck, P., and Xie, Y. (2021b).
\newblock Spatio-temporal point processes with attention for traffic congestion
  event modeling.
\newblock {\em IEEE Transactions on Intelligent Transportation Systems},
  23(7):7298--7309.

\bibitem[Zhu et~al., 2021c]{zhu2021imitation}
Zhu, S., Li, S., Peng, Z., and Xie, Y. (2021c).
\newblock Imitation learning of neural spatio-temporal point processes.
\newblock {\em IEEE Transactions on Knowledge and Data Engineering}.

\bibitem[Zhu et~al., 2022b]{zhu2022neural}
Zhu, S., Wang, H., Dong, Z., Cheng, X., and Xie, Y. (2022b).
\newblock Neural spectral marked point processes.
\newblock In {\em International Conference on Learning Representations}.

\bibitem[Zhu et~al., 2021d]{zhu2021deep}
Zhu, S., Zhang, M., Ding, R., and Xie, Y. (2021d).
\newblock Deep fourier kernel for self-attentive point processes.
\newblock In {\em International Conference on Artificial Intelligence and
  Statistics}, pages 856--864. PMLR.

\end{thebibliography}

\newpage


\newpage
\appendix

\section{Experimental setup and additional results}
\label{append:additional-experiments}

In this section, we first provide a detailed description of three baseline models used in Section \ref{sec:results} and their hyper-parameter selection. Then we present additional numerical results, including uncertainty quantification and comparison between models using different temporal kernels.

\subsection{Baseline description}

Homogeneous Poisson process assumes events occur at a constant intensity $\lambda$ over space. The parameter indicates the expected number of events occurring in a unit interval or region. We estimate $\lambda$ using the average number of confirmed cases over time and space and randomly sample events in the spatio-temporal space. The results of the homogeneous Poisson process act as a sanity check.

Susceptible-Infectious-Recovered (SIR) model is one of the most fundamental compartmental models that aim to model infectious disease spread. It splits the whole population into three compartments of susceptible ($\bm{S}$), infected ($\bm{I}$) and recovered ($\bm{R}$) individuals. SIR makes each compartment a function of $t$ since the population of each compartment may vary over time, and three ordinary differential equations about these three functions can describe the integral SIR system.
Parameters $\beta_{\text{SIR}}$ and $\gamma_{\text{SIR}}$ represent the emerging rate of new infections and the recovery rate of patients, respectively. Both parameters are fitted according to the real data based on least squares. We fit a SIR model for each comuna, choosing the initial infected population $\bm{I}(0)$ to be the number of cases at the week of the first case.

\begin{figure}[!ht]
\centering 
\includegraphics[width=.8\linewidth]{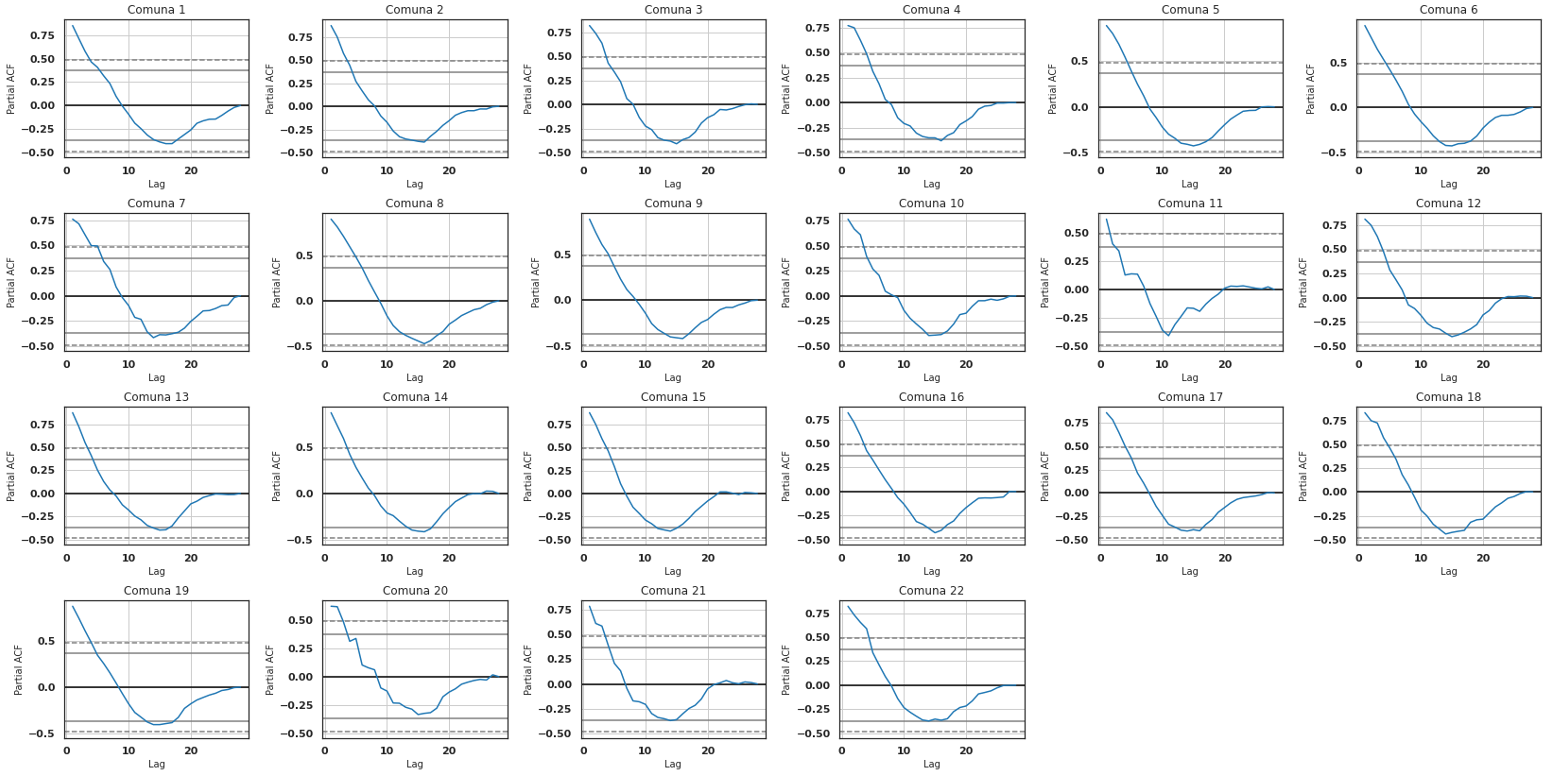}
\caption{PACF plot for each comuna. The $x$-axis is the lag number of the time series itself to the current output and the $y$-axis is the value of PACF at the corresponding lag. For example, the PACF of a time series $\bm{X}$ at lag 2 refers to the partial autocorrelation between $\bm{X}_{t}$ and $\bm{X}_{t-2}$. The dashed line represents the lower bound of significant PACF value.
}
\label{fig:PACF}
\end{figure}

Linear prediction is another popular method used to do forecasting tasks. We choose an autoregressive (AR) time series model to predict the number of infected cases. It specifies that the current output value depends linearly on its history and a white noise term. A parameter $p$ in AR model represents the number of most recent lags that the current output depends on, which can be determined by choosing the appropriate number of significant lags of PACF about the data series. We choose $p=3$ for the AR model according to PACF plots of confirmed case series in each comuna in Fig.~\ref{fig:PACF}.

ETAS is a benchmark model in modeling specific spatio-temporal data, as we mentioned in Section \ref{sec:proposed-method}. We replace the spatio-temporal kernel in \eqref{eq:general-form} with a Gaussian diffusion kernel. We estimate model parameters by applying stochastic gradient descent with regard to model likelihood.

\subsection{Uncertainty quantification of point process prediction}

\begin{figure}[!t]
    \centering 
    \includegraphics[width=.95\textwidth, height=2.8in]{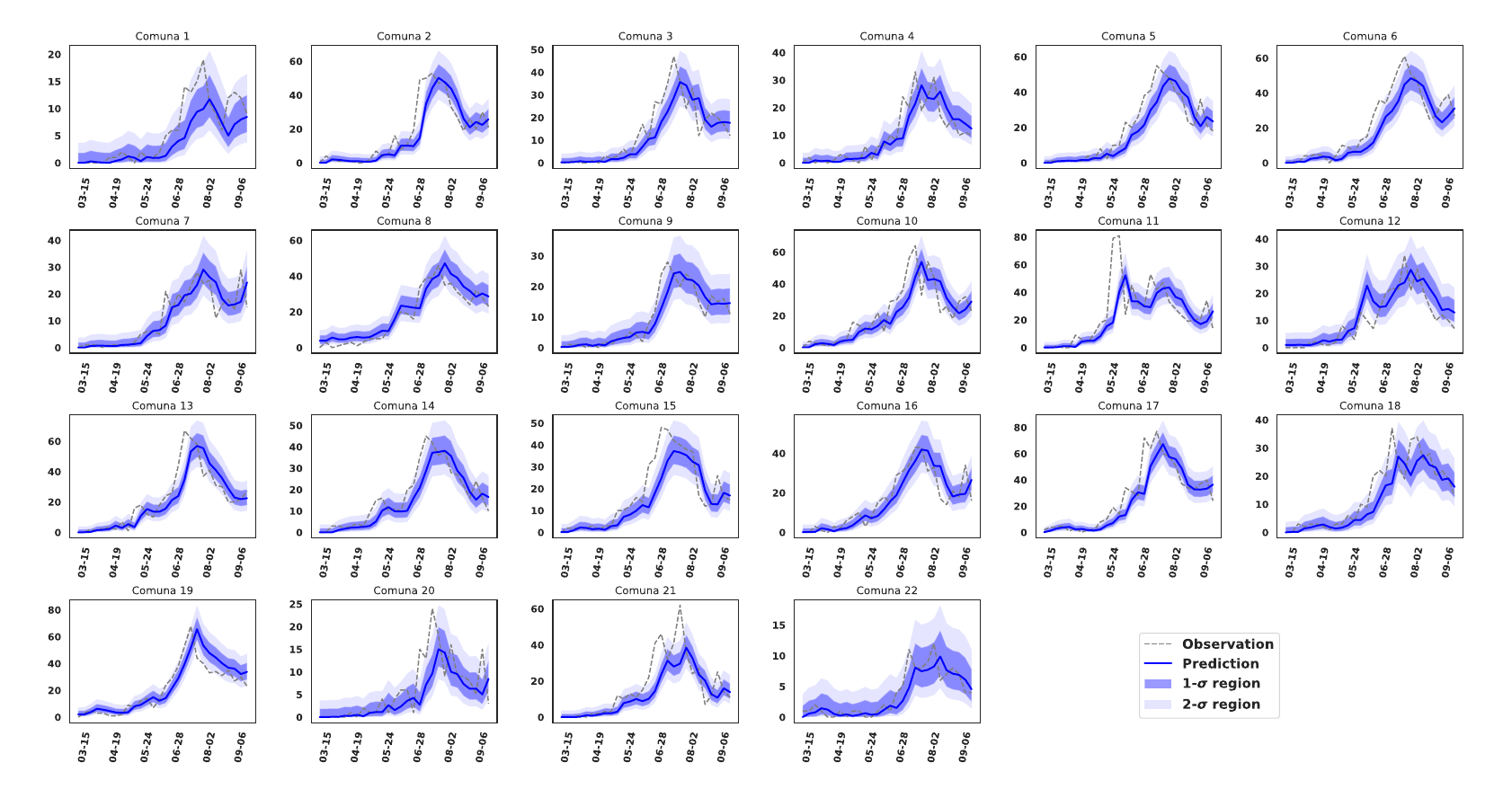}
    \caption{{\color{black} Uncertainty quantification of in-sample estimation. Gray dash lines represent the real case number. Blue lines indicate the mean of in-sample estimation. Shaded areas represent the confidence interval of the in-sample estimation, where different color depth indicates $1\sigma$ (68\%) and $2\sigma$ (95\%) regions, respectively.}}
    \label{fig:linechart-uq}
\end{figure}

We can quantify the uncertainty in the point process based prediction, since the number of cases in an arbitrary region $S$ is a non-homogeneous Poisson random variable. Specifically, according to \cite{Sahai1993UQ}, the confidence interval of one-week-ahead prediction $\hat\lambda$ using our method is given by
\[
        \hat\lambda_{l} = \frac{1}{2}\chi^2_{2\mu, \alpha/2},\quad\hat\lambda_{u} = \frac{1}{2}\chi^2_{2(\mu+1), 1-\alpha/2}
\]
where $\hat\lambda_{l}$ and $\hat\lambda_{u}$ denote the lower and upper bounds of the prediction, respectively. Note that the $\mu=\int_{S}\hat\lambda(t)$ is the predicted number of cases that occurred in the region $S$ at time $t$ and $\alpha$ is the confidence level. 
We include additional experimental results with uncertainty quantification in our work. The confidence intervals of predictions of our model are shown in Figure~\ref{fig:linechart-uq}. 
More quantitative results about uncertainty quantification for in-sample and out-of-sample predictions are summarized in Tables 1, 2, 3, and 4. Numbers in the brackets represent the standard deviation of predictions.

\subsection{Performance comparison of different temporal kernels}

\begin{figure}[!t]
\centering
\includegraphics[width=.95\textwidth, height=2.8in]{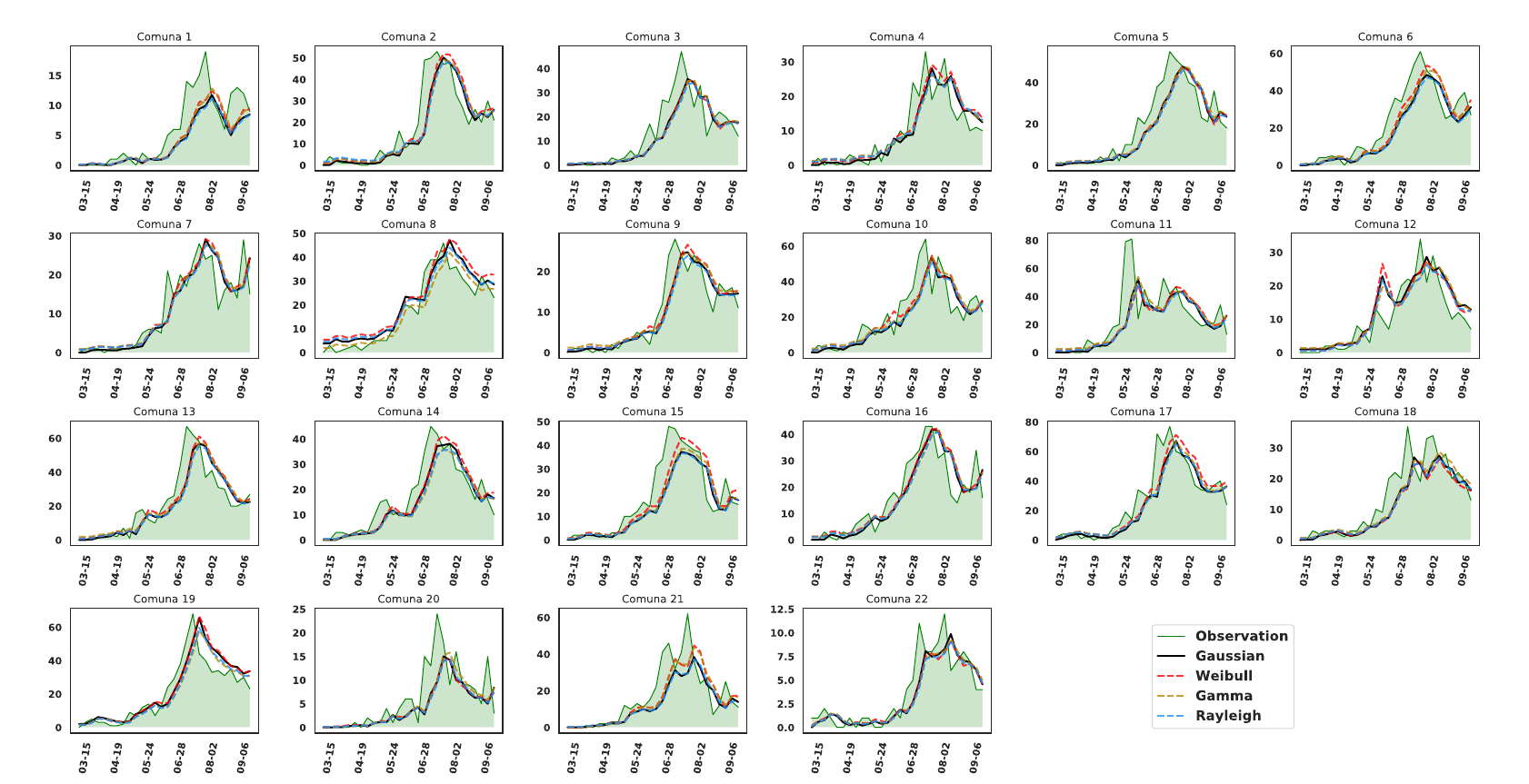}
\caption{{\color{black}Comparison of in-sample estimation using different temporal kernels. Solid green lines represent the observation of case numbers in each comuna. Solid black lines indicate the in-sample estimation using a Gaussian kernel. Red, yellow, and blue lines represent the in-sample estimation using Weibull, Gamma, and Rayleigh kernels, respectively.}}
\label{fig:temp-kernel-linechart-comparison}
\end{figure}


To further illustrate the effectiveness of our choice of the temporal kernel, we present the following ablation studies comparing the Gaussian kernel with three other temporal kernel functions:
\begin{itemize}
    \item Weibull kernel: $\nu(x) = \frac{k}{\lambda}\left(\frac{x}{\lambda}\right)^{k-1}e^{-(x/\lambda)^k}, x \geq 0$
    \item Gamma kernel: $\nu(x) = \frac{\beta^\alpha}{\Gamma(\alpha)}x^{\alpha-1}e^{-\beta x}, x \geq 0$
    \item Rayleigh kernel: $\nu(x) = \frac{x}{\sigma^2}e^{-x^2/(2\sigma^2)}, x \geq 0$
\end{itemize}
The in-sample estimation results are summarized in Figure~\ref{fig:temp-kernel-linechart-comparison} and Table 3. The results validate that the Gaussian temporal kernel marginally outperforms other candidates on real data by providing the lowest in-sample estimation error and variance.
The out-of-sample prediction results for the last four weeks are presented in Table 4, which also validates the predictive power of NSSTPP with the Gaussian kernel.

\begin{table}
\caption{{\color{black} Performance of in-sample estimation with different temporal kernels. The bold values mark the best performance among all models regarding different metrics.}}
\centering
\begin{tabular}{lcccc}
\hline
\hline
Models & Log-likelihood($\times 10^4$) & MAE $Q^\text{in}_{0.25}$ & MAE $Q^\text{in}_{0.5}$ & MAE $Q^\text{in}_{0.75}$ \\
\hline
NSSTPP($R$=3)+Weibull & $9.671_{(0.0369)}$ & $0.950_{(0.094)}$ & $2.738_{(0.177)}$ & $7.009_{(0.341)}$\\
NSSTPP($R$=3)+Gamma & $9.669_{(0.0364)}$ & $0.966_{(0.091)}$ & $2.777_{(0.171)}$ & $6.941_{(0.351)}$\\
NSSTPP($R$=3)+Rayleigh & $9.706_{(0.1208)}$ & $1.039_{(0.092)}$ & $2.946_{(0.177)}$ & $7.465_{(0.342)}$\\
NSSTPP($R$=3)+Gaussian & $9.331_{(0.0937)}$ & $\textbf{0.797}_{(0.085)}$ & $\textbf{2.620}_{(0.161)}$ & $\textbf{6.757}_{(0.330)}$\\
\hline  
\hline
\label{tab:temp-kernel-in-sample-estimation}
\end{tabular}
\end{table}

\begin{table}
\caption{{\color{black}Performance of out-of-sample estimation with different temporal kernels for last four week. The bold values mark the best performance among all models regarding different metrics.}
}
\centering
\resizebox{0.9\linewidth}{!}{
\begin{tabular}{lcccc}
\hline
\hline
Models & MAE $Q^\text{out}_{0.25}$ & MAE $Q^\text{out}_{0.5}$ & MAE $Q^\text{out}_{0.75}$ \\
\hline
NSSTPP($R$=3)+Weibull & $3.139_{(0.593)}$ & $6.673_{(0.777)}$ & $10.727_{(0.980)}$ \\
NSSTPP($R$=3)+Gamma & $2.853_{(0.590)}$ & $6.490_{(0.737)}$ & $11.013_{(0.971)}$ \\
NSSTPP($R$=3)+Rayleigh & $3.177_{(0.587)}$ & $6.521_{(0.765)}$ & $\textbf{10.212}_{(0.991)}$ \\
NSSTPP($R$=3)+Gaussian & $\textbf{2.843}_{(0.585)}$ & $\textbf{6.398}_{(0.751)}$ & $10.495_{(0.892)}$ \\
\hline  
\hline
\label{tab:temp-kernel-out-of-sample-estimation}
\end{tabular}
}
\end{table}

However, the form of the temporal kernel may well depend on specific tasks. The choice of the temporal kernel can be more systematically explored in other scenarios. Understanding and characterizing what kernels can be learned through such an approach is left for future study.


\section{Derivation of the point process log-likelihood}
\label{append:derivation-log-likelihood}

Assume that we have total number of $\mathbb{N}([0, T] \times \mathcal{S})$ observations in $\bm{x}$. For any given $t \in [0, T]$, we assume that $n$ events happened before $t$ and denote the occurrence time of the latest event as $t_n$. Let $\Omega = [t, t + dt) \times B(s, ds)$ where $s \in \mathcal{S}$. Let $F(t) = \mathbb{P}(\bm{x}_{n+1}, t_{n+1} < t | \mathcal{H}_{t_n} \cup \bm{x}_n)$ be the conditional cumulative probability function, and $\mathcal{H}_{t_n} \cup \bm{x}_n$ represents the history events happened up to time $t_n$ and at $t_n$. Let $f(t, s) \triangleq f(t, s|\mathcal{H}_{t_n} \cup \bm{x}_n)$ be the corresponding conditional probability density function of new event happening in $\Omega$. As defined in \eqref{eq:original-definition}, $\lambda(t, s)$ can be expressed as
\[
    \begin{aligned}
        \lambda(t, s) &= \mathbb{P}\{\bm{x}_{n+1} \in \Omega | \mathcal{H}_{t}\} = \mathbb{P}\{\bm{x}_{n+1} \in \Omega | \mathcal{H}_{t_n} \cup \bm{x}_n \cup \{t_{n+1} \geq t\}\} \\&= \frac{\mathbb{P}\{\bm{x}_{n+1} \in \Omega, t_{n+1} \geq t | \mathcal{H}_{t_{n}} \cup \bm{x}_{n}\}}{\mathbb{P}\{t_{n+1} \geq t | \mathcal{H}_{t_{n}} \cup \bm{x}_{n}\}} \\&= \frac{f(t, s)}{1 - F(t)}
    \end{aligned} 
\]
We multiply the differential of time and space $dtds$ on both side of the equation, and integral over $s$
\[
    dt \cdot \int_{\mathcal{S}}\lambda(t, u)du = \frac{dt \cdot \int_{\mathcal{S}}f(t, u)du}{1 - F(t)} = \frac{dF(t)}{1 - F(t)} = -d\log{(1 - F(t))}.
\]
Hence, integrating over $t$ on $(t_n, t)$ leads to $F(t) = 1 - \exp (-\int_{t_{n}}^{t}\int_{\mathcal{S}} \lambda(\tau, u)dud\tau)$ because $F(t_{n}) = 0$. Then we have
\begin{equation}
    f(t, s) = \lambda(t, s) \cdot \exp \left( -\int_{t_{n}}^{t}\int_{\mathcal{S}} \lambda(\tau, u)dud\tau \right)\ ,
    \nonumber
\end{equation}
The joint p.d.f. for a realization is then, by the chain rule, $f(x_1, ..., x_{\mathbb{N}([0, T]\times \mathcal{S})}) = \prod_{i=1}^{\mathbb{N}([0, T]\times \mathcal{S})} f(t_i, s_i)$. Then the log-likelihood of an observed sequence $\bm{x}$ can be written as
\[
    l(\bm{x}) = \sum_{i=1}^{\mathbb{N}([0, T]\times \mathcal{S})} \log \lambda(t_i, s_i) - \int_{0}^{T}\int_{\mathcal{S}} \lambda(\tau, u)dud\tau .
\]

\section{Derivation of the non-stationary spatial kernel}
\label{append:spatial-kernel-derivation}

In this section, we prove the formulation of the function $v(s, s^\prime)$ between two bivariate normal kernels as appears in \eqref{eq:spatial-kernel-expression-one-component}. Let two independent bivariate Gaussian variables $X_s, X_{s^\prime}$ be centered at locations $s, s^\prime$ with $\Sigma_s, \Sigma_{s^\prime}$ parameterized by
\[
\Sigma_s = \begin{pmatrix} a^2 & \rho a b \\ \rho a b & b^2 \end{pmatrix}, \ \Sigma_{s^{'}} = \begin{pmatrix} {a^\prime}^2 & \rho^\prime a^\prime b^\prime \\ \rho^\prime a^\prime b^\prime & {b^\prime}^2 \end{pmatrix},
\]
By common knowledge, the probability density function $f_Z$ of the sum $Z$ of two independent random variables $X, Y$, i.e $Z = X + Y$, is the convolution of the probability density functions $f_X$ and $f_Y$, i.e.
\[
f_Z(z) = \int_{-\infty}^{\infty} f_Y(z - x)f_X(x)dx,
\]
In our case, let us denote the probability density function of $X_s, X_{s^\prime}$ as $\kappa_s(\cdot), \kappa_{s^\prime}(\cdot)$. Then, we have
\[
f_{X_s+X_{s^\prime}}(x) = \int_{\mathbb{R}^2} \kappa_s(u)\kappa_{s^\prime}(x-u)du,
\]
We also have the following equalities due to the property of Gaussianity
\[
\kappa_{s}(2s-u) = \kappa_{s}(u), \kappa_{s^\prime}(2s^\prime-u) = \kappa_{s^\prime}(u).
\]
Writing $x = 2s^\prime$, we therefore have
\[
f_{X_s+X_{s^\prime}}(2s^\prime) = \int_{\mathbb{R}^2} \kappa_s(u)\kappa_{s^\prime}(u)du = v(s, s^\prime),
\]
Now it is easy to write that $X_s+X_{s^\prime} \sim \mathcal{N}\left ( s + s^\prime, \Sigma_s + \Sigma_{s^\prime} \right )$, and thus
\[
    \begin{aligned}
         v(s, s^\prime) &= f_{X_s+X_{s^\prime}}(2s^\prime) = \frac{1}{2\pi | \Sigma_s + \Sigma_{s^\prime} |^{\frac{1}{2}}} \exp\left \{-\frac{1}{2}(s^\prime - s)^\top(\Sigma_s + \Sigma_{s^\prime})^{-1}(s^\prime - s)\right \} \\ &=\frac{1}{q_1} \exp\left \{-\frac{1}{q_2}(s - s^\prime)^\top W (s - s^\prime)\right \},
    \end{aligned}
\]
where
\[
    \begin{aligned}
    W &= \begin{pmatrix} b^2 + {b^\prime}^2 & -(\rho a b + \rho^\prime a^\prime b^\prime) \\ -(\rho a b + \rho^\prime a^\prime b^\prime) & a^2 + {a^\prime}^2 \end{pmatrix}, \\
    q_1 &= 2\pi | \Sigma_s + \Sigma_{s^\prime} |^{\frac{1}{2}} \\
    &= 2\pi \sqrt{-(2\rho \rho^\prime a a^\prime b b^\prime + a^2((\rho^2 -1)b^2 - {b^\prime}^2) + {a^\prime}^2(({\rho^\prime}^2 - 1){b^\prime}^2 - b^2))}, \\
    q_2 &= -2(2\rho \rho^\prime a a^\prime b b^\prime + a^2((\rho^2 -1)b^2 - {b^\prime}^2) + {a^\prime}^2(({\rho^\prime}^2 - 1){b^\prime}^2 - b^2)). \quad
    \end{aligned}
\]

\section{Derivation of the covariance function}
\label{append:covariance-derivation}

Assume an ellipse centered at the origin with area $A$ and two focus points of the ellipse in $\mathbb{R}^2$ which are $(\bm{\psi}_x, \bm{\psi}_y), (-\bm{\psi}_x, -\bm{\psi}_y)$, where $\bm{\psi}_x, \bm{\psi}_y \in \mathbb{R}$. In what follows, we use the same notation of $\Sigma$ as before. For the ellipse parameters, we denote the semi-major and semi-minor axis of the ellipse as $\sigma_1, \sigma_2$. According to the ellipse formula, we have
\[
    \left \{ 
    \begin{array}{rl}
        \pi \sigma_1 \sigma_2 &= A \\
        \sigma_1^2 - \sigma_2^2 &= \bm{\psi}_x^2 + \bm{\psi}_y^2 = \|\bm{\psi}\|^2
    \end{array}
    \right .\ .
\]
We can also compute 
\begin{equation}
    \sigma_1 = \left(\frac{\sqrt{4A^2 + \|\bm{\psi}\|^4\pi^2}}{2\pi} + \frac{\|\bm{\psi}\|^2}{2}\right)^{\frac{1}{2}}, \quad
    \sigma_2 = \left(\frac{\sqrt{4A^2 + \|\bm{\psi}\|^4\pi^2}}{2\pi} - \frac{\|\bm{\psi}\|^2}{2}\right)^{\frac{1}{2}}.
    \label{Ellipse}
\end{equation}
As the rotation angle $\alpha$ of the ellipse is $\alpha = \tan^{-1}(\bm{\psi}_y/\bm{\psi}_x)$, we can write the bivariate normal random variable $X$ as follows
\[
X = \begin{pmatrix} \cos{\alpha} & -\sin{\alpha} \\ \sin{\alpha} & \cos{\alpha} \end{pmatrix} Z,
\]
where $Z = \begin{pmatrix} Z_1 \\ Z_2 \end{pmatrix}$ with covariance matrix $\begin{pmatrix} \sigma_1^2 & 0 \\ 0 & \sigma_2^2 \end{pmatrix}$. 

Now we introduce the kernel scale parameter $\tau_z$, and write down the covariance matrix of $X$ as
\[
    \Sigma = \tau_z^2 \begin{pmatrix} \sigma_1^2\cos^2 \alpha + \sigma_2^2 \sin^2\alpha & (\sigma_1^2 - \sigma_2^2)\cos{\alpha}\sin{\alpha} \\ (\sigma_1^2 - \sigma_2^2)\cos{\alpha}\sin{\alpha} & \sigma_1^2\sin^2 \alpha + \sigma_2^2\cos^2\alpha \end{pmatrix}\ .
\]
Plugging \eqref{Ellipse} into this equation we get
\begin{equation}
    \Sigma_s = \tau_z^2 \begin{pmatrix} Q + \frac{\|\bm{\psi}\|^2}{2}\cos2\alpha & \frac{\|\bm{\psi}\|^2}{2}\sin2\alpha \\ \frac{\|\bm{\psi}\|^2}{2}\sin2\alpha & Q - \frac{\|\bm{\psi}\|^2}{2}\cos2\alpha \end{pmatrix},
\end{equation}
where $Q = \sqrt{4A^2 + \|\bm{\psi}\|^4\pi^2}/2\pi, \alpha = \tan^{-1}(\bm{\psi}_y / \bm{\psi}_x)$.

\section{Proof of Section~\ref{sec:efficient-learning}}
\label{append:proof-efficient-learning}

For convenience, we denote our approximation of the intractable double integral, $\lambda_0|\mathcal{S}|T + T \sum_{l=1}^{L}\gamma_l + \sqrt{2\pi}C\sigma_0 \sum_{i=1}^{\mathbb{N}([0, T] \times \mathcal{S})}\left\{h\left(\frac{T - t_i}{\sigma_0}\right) - \frac{1}{2} \right\}$, as $I_{\text{Approx}}$. The boundary effect error $\epsilon_1$ satisfies that $\int_0^T\int_{\mathcal{S}}\lambda(\tau, r)drd\tau + \epsilon_1 = \int_0^T\int_{\mathbb{R}^2}\lambda(\tau, r)drd\tau$ according to the first assumption. Based on the second assumption, we approximate the location-dependent kernel induced feature functions $\kappa_s$ with $\kappa_{s}^{0}$ to solve the intractable integration in \eqref{eq:integral-term}. We first derive the upper bound $\eta_{\mathrm{bound}}(A, c)$ of the approximation error $\left < \kappa_s^{(r_1)}, \kappa_{s^\prime}^{(r_2)} \right > - \left < \kappa_{s}^{0}, \kappa_{s^\prime}^{(r_2)} \right >$. Given any location $s$ and history event $s^\prime$, for the convenience of computation while without loss of generality, we can locate the origin of the coordinate system at $s$ and align $x$-axis and $y$-axis with the semi-major and semi-minor axis of the one standard ellipse of $\kappa_s$. Thus according to Section \ref{append:spatial-kernel-derivation}, the second inner-product $\left < \kappa_{s}^{0}, \kappa_{s^\prime}^{(r_2)} \right >$ equals to the probability density function $f_{X_0 + X_{s^\prime}^{(r_2)}}(2(s^\prime - s))$ of the summation of two independent random variable $X_0$ and $X_{s^\prime}^{(r_2)}$, where $X_0 \sim \mathcal{N}(\bm{0}, \Sigma_0), \Sigma_0 = \frac{\tau_z^2A}{\pi}\mathbf{I}$ and $X_{s^\prime}^{(r_2)} \sim \mathcal{N}(s^\prime - s, \Sigma_{s^\prime}^{(r_2)})$, $\Sigma_{s^\prime}^{(r_2)}$ is the covariance matrix of $\kappa_{s^\prime}^{(r_2)}$ in the preset coordinate system.
For the first term we re-write the inner-product in polar coordinate system and have
\begin{align*}
    \left < \kappa_s^{(r_1)}, \kappa_{s^\prime}^{(r_2)} \right > &= \int_{\mathbb{R}^2}\kappa_s^{(r_1)}(u)\kappa_{s^\prime}^{(r_2)}(u)du \\&= \int_0^{2\pi} \int_0^{+\infty} \kappa_s^{(r_1)}(r, \theta)\kappa_{s^\prime}^{(r_2)}(r, \theta) r dr d\theta \\&\overset{\text{(i)}}{=} \int_0^{2\pi} \int_0^{+\infty} \frac{r}{2\pi\tau_z^2\sqrt{Q^2 - \frac{\|\bm{\psi}_s\|^4}{4}}}\exp{\left\{-\frac{r^2}{2\tau_z^2}\left(\frac{\cos^2{\theta}}{Q + \frac{\|\bm{\psi}_s\|^2}{2}} + \frac{\sin^2\theta}{Q - \frac{\|\bm{\psi}_s\|^2}{2}}\right)\right\}} \\ &\quad \cdot \kappa_{s^\prime}^{(r_2)}(r, \theta)drd\theta \\&= \int_0^{2\pi} \int_0^{+\infty} \frac{r}{2\tau_z^2A}\exp{\left\{-\frac{r^2\pi^2}{2\tau_z^2A^2}\left(Q - \frac{\|\bm{\psi}_s\|^2}{2}\cos2\theta\right)\right\}}  \cdot \kappa_{s^\prime}^{(r_2)}(r, \theta)drd\theta \\&\overset{\text{(ii)}}{\leq} \int_0^{2\pi} \int_0^{+\infty} \frac{r}{2\tau_z^2A}\exp{\left\{-\frac{r^2\pi}{\tau_z^2(\sqrt{4A^2 + c^4\pi^2} + c^2\pi)}\right\}} \cdot \kappa_{s^\prime}^{(r_2)}(r, \theta)drd\theta, \quad \forall r_1, r_2. \numberthis \label{eq:upper-bound-scaling}
\end{align*}
Here $Q$ is defined in Append \ref{append:covariance-derivation}. We plug in the analytical form of $\kappa_s^{(r_1)}(r, \theta)$ at (i). The inequality at (ii) holds because:
\begin{align*}
    Q - \frac{\|\bm{\psi}_s\|^2}{2}\cos2\theta &\geq Q - \frac{\|\bm{\psi}_s\|^2}{2} = \frac{\sqrt{4A^2 + \|\bm{\psi}_s\|^4\pi^2} - \|\bm{\psi}_s\|^2\pi}{2\pi} \\&= \frac{2A^2}{\pi(\sqrt{4A^2 + \|\bm{\psi}_s\|^4\pi^2} + \|\bm{\psi}_s\|^2\pi)} \\&\geq \frac{2A^2}{\pi(\sqrt{4A^2 + c^4\pi^2} + c^2\pi)} \quad (\text{because} \ \|\bm{\psi}_s\| \leq c) \ .
\end{align*}
For the final formula on the right of the inequality operator, we find that
\[
    \begin{aligned}
        &\quad \frac{1}{2\tau_z^2A}\exp{\left\{-\frac{r^2\pi}{\tau_z^2(\sqrt{4A^2 + c^4\pi^2} + c^2\pi)}\right\}} \\ &= \frac{\sqrt{4A^2 + c^4\pi^2} + c^2\pi}{2A} \cdot \frac{1}{2\pi \frac{\tau_z^2}{2}\left(\sqrt{4A^2/\pi^2 + c^4} + c^2\right)}\exp{\left\{-\frac{r^2}{2\frac{\tau_z^2}{2}\left(\sqrt{4A^2/\pi^2 + c^4} + c^2\right)}\right\}} \ .
    \end{aligned}
\]
It takes the form of the multiplication of a constant and a probability density function of a Gaussian distribution with zero mean and covariance matrix $\Sigma_1$, where $\Sigma_1 = \tau_z^2\frac{\sqrt{4A^2/\pi^2 + c^4} + c^2}{2}\mathbf{I}$.
We assume a random variable $X_1$ conforms to the corresponding distribution $X_1 \sim \mathcal{N}(\bm{0}, \Sigma_1)$ and denote the constant $\frac{\sqrt{4A^2 + c^4\pi^2} + c^2\pi}{2A}$ as $U$ (the same one in the proposition \ref{prop:approximation-of-integral}). Combining the result of \eqref{eq:upper-bound-scaling} we can write the upper bound of the approximation error as
\begin{equation}    
    \left < \kappa_s^{(r_1)}, \kappa_{s^\prime}^{(r_2)} \right > - \left < \kappa_{s}^{0}, \kappa_{s^\prime}^{(r_2)} \right > \leq U \cdot f_{X_1 + X_{s^\prime}^{(r_2)}}(2(s^\prime - s)) - f_{X_0 + X_{s^\prime}^{(r_2)}}(2(s^\prime - s)) \ .
    \label{eq:approxiamtion-upper-bound}
\end{equation}
We introduce a new notation $\upsilon_0(s, s^\prime)$ to denote the spatial kernel with $\kappa_s$ replaced with $\kappa_{s}^{0}$, that is $\upsilon_0(s, s^\prime) = \sum_{(r_1, r_2) \in [R] \times [R]} w_s^{(r_1)}w_{s_i}^{(r_2)} \left< \kappa_{s}^{0}, \kappa_{s_i}^{(r_2)} \right>, [R] = \{1, 2, ..., R\}$. Based on above results, for any given history event $s_i$ we have
\begin{align*}
    \\&\quad \int_{\mathbb{R}^2}\upsilon(u, s_i)du - \int_{\mathbb{R}^2}\upsilon_0(u, s_i)du \\&= \int_{\mathbb{R}^2} \sum_{(r_1, r_2) \in [R] \times [R]} w_u^{(r_1)} w_{s_i}^{(r_2)} \left( \left< \kappa_{u}, \kappa_{s_i}^{(r_2)} \right> - \left< \kappa_{u}^{0}, \kappa_{s_i}^{(r_2)} \right> \right)du \\&\leq \int_{\mathbb{R}^2} \sum_{(r_1, r_2) \in [R] \times [R]} w_u^{(r_1)} w_{s_i}^{(r_2)} \left(U\cdot f_{X_1 + X_{s_i}^{(r_2)}}(2(s_i - u)) - f_{X_0 + X_{s^\prime}^{(r_2)}}(2(s^\prime - u)) \right)du \quad (\text{Plug in}\ \eqref{eq:approxiamtion-upper-bound}) \\&= \int_{\mathbb{R}^2} \sum_{r_2 = 1}^{R} w_{s_i}^{(r_2)} \left(U \cdot f_{X_1 + X_{s^\prime}^{(r_2)}}(2(s^\prime - u)) - f_{X_0 + X_{s^\prime}^{(r_2)}}(2(s^\prime - u)) \right)du \quad (\text{sum over} \ r_1) \\&= \sum_{r_2 = 1}^{R} w_{s_i}^{(r_2)} \left( U \int_{\mathbb{R}^2}f_{X_1 + X_{s^\prime}^{(r_2)}}(2(s^\prime - u))du - \int_{\mathbb{R}^2}f_{X_0 + X_{s^\prime}^{(r_2)}}(2(s^\prime - u))du \right) \\&= \sum_{r_2 = 1}^{R} w_{s_i}^{(r_2)} \left(U - 1\right) (\text{Integration of probability density function over}\ \mathbb{R}^2\ \text{equals to}\ 1) \\&= U - 1 \ .
\end{align*}
Also notice that
\begin{align*}
    \int_{\mathbb{R}^2}\upsilon_0(u, s_i)du &= \int_{\mathbb{R}^2}\sum_{(r_1, r_2) \in [R] \times [R]} w_u^{(r_1)}w_{s_i}^{(r_2)} \left < \kappa_{s}^{0}, \kappa_{s_i}^{(r_2)} \right >du \\ &= \int_{\mathbb{R}^2} \sum_{r_2=1}^{R}w_{s_i}^{(r_2)}\left < \kappa_{s}^{0}, \kappa_{s_i}^{(r_2)} \right >du \quad (\mathrm{because} \sum_{r_1 = 1}^{R}w_u^{(r_1)} = 1) \\ &= \sum_{r_2=1}^{R}w_{s_i}^{(r_2)} \int_{\mathbb{R}^2}\left < \kappa_{s}^{0}, \kappa_{s_i}^{(r_2)} \right >du = \sum_{r_2=1}^{R}w_{s_i}^{(r_2)} = 1 \ .
\end{align*}
Thus the upper bound of $\epsilon_2$ can be controlled by
\begin{align*}
    \epsilon_2 &= \left( \int_0^T\int_{\mathbb{R}^2}\lambda(\tau, r)drd\tau - I_{\text{Approx}} \right) \bigg/ I_{\text{Approx}} \\ &= \left\{ \lambda_0|\mathcal{S}|T + T \sum_{l=1}^{L}\gamma_l + \int_{0}^{T} \sum_{t_i < \tau} Ce^{-\frac{1}{2\sigma_0^2}(\tau-t_i)^2}d\tau \cdot \int_{\mathbb{R}^2} \upsilon(u, s_i)du - I_{\text{Approx}} \right\} \bigg/ I_{\text{Approx}} \\ &= \left\{ \int_0^{T} \sum_{t_i < \tau} Ce^{-\frac{1}{2\sigma_0^2}(\tau-t_i)^2}d\tau \left(\int_{\mathbb{R}^2} \upsilon(u, s_i)du - 1\right) \right\} \bigg/ I_{\text{Approx}} \\ &= \left\{ \int_0^{T} \sum_{t_i < \tau} Ce^{-\frac{1}{2\sigma_0^2}(\tau-t_i)^2}d\tau \int_{\mathbb{R}^2} \left( \upsilon(u, s_i) - \upsilon_0(u, s_i) \right)du \right\} \bigg/ I_{\text{Approx}} \\ & \leq \left\{\left( U - 1 \right) \int_0^{T} \sum_{t_i < \tau} Ce^{-\frac{1}{2\sigma_0^2}(\tau-t_i)^2}d\tau \right\} \bigg/ I_{\text{Approx}} \ ,
\end{align*}
because $I_{\text{Approx}} > \int_0^{T} \sum_{t_i < \tau} Ce^{-\frac{1}{2\sigma_0^2}(\tau-t_i)^2}d\tau = \sqrt{2\pi}C\sigma_0 \sum_{i=1}^{\mathbb{N}([0, T] \times \mathcal{S})}\left\{h\left(\frac{T - t_i}{\sigma_0}\right) - \frac{1}{2} \right\}$, we have
\begin{align*}
    \epsilon_2 < U - 1 \ . \numberthis \label{eq:final-upper-bound}
\end{align*}
On the other hand,
\begin{align*}
    \left < \kappa_s^{(r_1)}, \kappa_{s^\prime}^{(r_2)} \right > &= \int_0^{2\pi} \int_0^{+\infty} \frac{r}{2\tau_z^2A}\exp{\left\{-\frac{r^2\pi^2}{2\tau_z^2A^2}\left(Q - \frac{\|\bm{\psi}_s\|^2}{2}\cos2\theta\right)\right\}}  \cdot \kappa_{s^\prime}^{(r_2)}(r, \theta)drd\theta \\&\overset{\text{(iii)}}{\geq} \int_0^{2\pi} \int_0^{+\infty} \frac{r}{2\tau_z^2A}\exp{\left\{-\frac{r^2\pi(\sqrt{4A^2 + c^4\pi^2} + c^2\pi)}{4\tau_z^2A^2}\right\}} \cdot \kappa_{s^\prime}^{(r_2)}(r, \theta)drd\theta \\&= \frac{1}{U}f_{X_2 + X_{s^\prime}^{(r_2)}}(2(s^\prime - s)) - f_{X_0 + X_{s^\prime}^{(r_2)}}(2(s^\prime - s)) \ , \numberthis \label{eq:approxiamtion-lower-bound}
\end{align*}
where $X_2 \sim \mathcal{N}(\bm{0}, \Sigma_2), \Sigma_2 = \frac{\tau_z^2A}{\pi U}\mathbf{I}$. The establishment of (iii) can be deduced by
\begin{align*}
    Q - \frac{\|\bm{\psi}_s\|^2}{2}\cos2\theta &\leq Q + \frac{\|\bm{\psi}_s\|^2}{2} \\&= \frac{\sqrt{4A^2 + \|\bm{\psi}_s\|^4\pi^2} + \|\bm{\psi}_s\|^2\pi}{2\pi} \\&\leq \frac{\sqrt{4A^2 + c^4\pi^2} + c^2\pi}{2\pi} \ .
\end{align*}
We can use the lower bound \eqref{eq:approxiamtion-lower-bound} to similarly obtain
\[
    \int_{\mathbb{R}^2}\upsilon(u, s_i)du - \int_{\mathbb{R}^2}\upsilon_0(u, s_i)du \geq \frac{1}{U} - 1 \ .
\]
Thus we have
\[
    \epsilon_2 \geq \left\{\left( \frac{1}{U} - 1 \right)  \int_0^{T} \sum_{t_i < \tau} Ce^{-\frac{1}{2\sigma_0^2}(\tau-t_i)^2}d\tau \right\} \bigg/ I_{\text{Approx}} \geq \frac{1}{U} - 1 \ . \ (\text{because} \ \frac{1}{U} - 1 < 0)
\]
Combining the result \eqref{eq:final-upper-bound} we have
$$
|\epsilon_2| < \max \left \{ U - 1, 1 - \frac{1}{U} \right \}. 
$$


\end{document}